\documentclass[%
reprint,
aps,
prb,
]{revtex4-2}
\usepackage{amsmath}
\usepackage{amssymb}
\usepackage{graphicx}
\usepackage{dcolumn}
\usepackage{bm}
\usepackage{amssymb}
\usepackage{physics}
\usepackage{xcolor}
\usepackage{multirow}
\usepackage{amsmath}
\usepackage{multirow}
\usepackage{soul}
\usepackage{upgreek}
\usepackage{makecell}
\begin{document} 

\newcommand{\transmonfreqsymb}{$\omega_\mathrm{q} /2 \pi$}
\newcommand{\storagelifetimesymb}{$T_{1,\mathrm{s}}$}
\newcommand{\transmonlifetimesymb}{$T_{1,\mathrm{q}}$}
\newcommand{\transmonechosymb}{$T_{2\mathrm{E},\mathrm{q}}$}
\newcommand{\transmonramseysymb}{$T_{2,\mathrm{q}}^*$}
\newcommand{\dispersiveshiftsymb}{$2\chi_{\mathrm{sq}}/2\pi$}
\newcommand{\seconddispshiftsymb}{$2\chi_{\mathrm{sq}} ^\prime /2\pi$}
\newcommand{\storagefreqsymb}{$\omega_\mathrm{s} /2 \pi$} 
\newcommand{\storagecoherencesymb}{$T_{2,\mathrm{s}}$}
\newcommand{\kerrsymb}{$K_{\mathrm{s}}/2\pi$}
\newcommand{\transmonlifetime}{72(3) $\upmu$s}                   %
\newcommand{\transmonramsey}{41(2) $\upmu$s}                     %

\newcommand{\resettotalduration}{800 ns}
\newcommand{\storagelifetime}{360(10) $\upmu$s}                   
\newcommand{\storagecoherencetime}{640(40) $\upmu$s}          
\newcommand{\transmonecho}{47(4) $\upmu$s}                       
\newcommand{\ecdtotaltime}{450 ns}                         
\newcommand{\fogifreq}{2.035833000 GHz}                     
                    
\newcommand{\dispersiveshift}{-45.44 kHz}                  
\newcommand{\kerr}{-12.854 Hz}                              
\newcommand{\seconddispshif}{4.190 Hz}                      
\newcommand{\kappar}{2.902 MHz}                             
\newcommand{\kappaf}{26.95 MHz}                           
\newcommand{\readouttime}{1 $\upmu$s}
\newcommand{\preechodelayInit}{400 ns}
\newcommand{\postechodelayInit}{120 ns}
\newcommand{\preechodelayRFE}{120 ns}
\newcommand{\postechodelayRFE}{1.24 $\upmu$s}

\newcommand{\displacementwidth}{50 ns}
\newcommand{\ecdinteractiontime}{100 ns}
\newcommand{\pipulsewidth}{10 ns}
\newcommand{\ecdfidelity}{$8(1)\times 10^{-3}$}
\newcommand{\readoutfreqsymb}{$\omega_\mathrm{r}/2\pi$}
\newcommand{\filterfreqsymb}{$\omega_\mathrm{f}/2\pi$}
\newcommand{\qubitresonatordispshiftsymb}{$\chi_\mathrm{qr}/2\pi$}
\newcommand{\qubitresonatordispshift}{-1.333 MHz}
\newcommand{\kapparsymb}{$\kappa_{1, \mathrm{r}}/2\pi$}
\newcommand{\kappafsymb}{$\kappa_\mathrm{f}^\mathrm{ext.}/2\pi$}
\newcommand{\qubitanharmsymb}{$K_\mathrm{q}/2\pi$}
\newcommand{\qubitanharm}{-0.203 GHz}
\newcommand{\timeofsBsround}{2.5 $\upmu$s}

\preprint{APS/123-QED}

\title{Quantum error correction of a grid-state qubit with state preparation and measurement errors below $\mathbf{ 10^{-3}}$}%

\author{Sara Turcotte$^{1, 2}$, Lucas St-Jean$^{1}$, Amélie L. Pessonneaux$^{1}$, Ross Shillito$^{1}$, \\ Bohdan Kulchytskyy$^{1}$, Eliott Ouellet$^{1}$, Jean Olivier Simoneau$^{1}$,  Florian Hopfmueller$^{1}$, Matthew Hamer$^{1}$, Pascal Lemieux$^{1}$, Dany Lachance-Quirion$^{1}$, Baptiste Royer$^{2}$, Nicholas E. Frattini$^{1}$}

\affiliation{%
 1. Nord Quantique, Sherbrooke, Québec J1K 1B7, Canada
}%
\affiliation{%
 2. Département de Physique et Institut quantique, Université de Sherbrooke, Québec, Canada
}%

\date{\today}%

\begin{abstract}
Grid state qubits offer a hardware-efficient approach to large-scale fault-tolerant quantum computing. They access the information redundancy required for quantum error correction by exploiting the large Hilbert space naturally available in harmonic oscillators. Superconducting architectures are particularly suitable to implement grid state qubits due to their fast and high-fidelity operations. Grid states in superconducting circuits enable quantum error correction (QEC) with performance beyond break-even. However, the state preparation and measurements (SPAM) errors of grid states has been a significant limitation to computational performances. In this work, we leverage high-performance QEC to enable repeat-until-success state preparation of both cardinal and magic states of the single-mode grid-state qubit. We combine this with an improved measurement protocol that corrects for both finite-energy envelope and auxiliary qubit readout errors, and increases robustness to photon loss. Our experiments, using both techniques, achieve a combined state-preparation and measurement error below $10^{-3}$. This represents two orders-of-magnitude improvement over the state of the art, bringing this platform on par with standard SPAM error levels measured in transmon qubits.
\end{abstract}

\maketitle

\section{\label{sec:intro}Introduction}

Quantum error correction is essential for the development of utility-scale quantum processors. Recent progress in bosonic quantum error correction has positioned this approach as a leading candidate for fault-tolerant quantum computation \cite{mirrahimi_dynamically_2014, michael_new_2016, albert_performance_2018, terhal_towards_2020, grimsmo_quantum_2021, lemonde_hardware-efficient_2024}. In these architectures, logical information is encoded in the large Hilbert space of high-quality oscillators, coupled to an auxiliary qubit enabling control and measurement operations. The intrinsic redundancy required for quantum error correction is provided by the oscillator Hilbert space, thus significantly reducing physical resource requirements, without introducing additional noise source. This makes bosonic codes a hardware-efficient route toward robust logical qubits.

Several bosonic encoding have been explored experimentally, including cat codes \cite{ofek_extending_2016, leghtas_confining_2015, touzard_coherent_2018, grimm_stabilization_2020, reglade_quantum_2024}, binomial codes \cite{hu_quantum_2019, ni_beating_2023}, dual-rail encodings \cite{teoh_dual-rail_2023, koottandavida_erasure_2024, levine_demonstrating_2024, chou_superconducting_2024, de_graaf_mid-circuit_2025} and grid codes \cite{campagne-ibarcq_quantum_2020, de_neeve_error_2022, sivak_real-time_2023, brock_quantum_2025, matsos_robust_2024, lachance-quirion_autonomous_2024,matsos_universal_2025}. Grid codes, also known as Gottesman–Kitaev–Preskill (GKP) codes \cite{gottesman_encoding_2001}, are particularly attractive due to their resilience to small displacement errors and single-photon loss. Circuit quantum electrodynamics (cQED) architectures provide an ideal platform for implementing GKP codes by combining high-coherence oscillator modes with fast and high-fidelity control. Moreover, the dominant error channel of these systems, single photon loss, naturally aligns with error-correcting properties of grid codes \cite{albert_performance_2018}. Recent experiments have demonstrated quantum error correction of single-mode GKP states beyond the break-even point, where the lifetime of the logical qubit exceeds that of its physical constituents \cite{sivak_real-time_2023, brock_quantum_2025}, and leveraged QEC to improve gate performance \cite{fontbote_beam_splitter}. This highlights the promise of bosonic architectures for scalable fault-tolerant quantum computation.

However, state preparation and measurement (SPAM) errors remain a major limitation to overall performance, as these errors are not directly addressed by standard error correction alone. In this work, we experimentally demonstrate in a circuit quantum electrodynamics architecture two distinct protocols, one dedicated to high-fidelity state preparation and the other to logical measurement of single-mode grid states, substantially reducing the total SPAM error.

For state preparation, we implement a postselected stabilization protocol that builds on high-performance quantum error correction \cite{campagne-ibarcq_quantum_2020}. We use a repeat-until-success compatible approach, which enables the state preparation of the six cardinal states of the single-mode grid code -- the eigenstates of the logical Pauli operators. Using a similar state preparation protocol, we also prepare $H$-type magic states with comparable fidelities \cite{baragiola_all-gaussian_2019}. These states serve as key resource to implement non-Clifford gates required for fault-tolerant quantum computation. 

We improve the logical measurements of grid-state qubits by applying an error-mitigation strategy in which finite-energy measurements are repeated within a single shot. Finite-energy measurements are the best known protocol to measure Pauli operators of grid codes, taking into account their finite-energy nature \cite{royer_stabilization_2020, hastrup_improved_2021, singh_towards_2025}. When combined with repeated measurements, this approach also suppresses the impact of errors from auxiliary qubit readout and adds robustness to photon loss errors, ultimately enhancing the overall fidelity of logical operator measurements.

When combined, state preparation through postselected stabilization and repeated finite-energy measurement can drastically reduce the total SPAM errors. Experimental results, leveraging these two strategies, demonstrate a total SPAM errors averaged over all cardinal states  below $7(7)\times 10^{-4}$ with a survival probability of 0.24 and 0.39 for state preparation and measurement respectively \footnote{Here and throughout, errors bars denotes a confidence interval of 95\%}. For magic states, the total SPAM error reaches $8(5)\times 10^{-3}$ with a total survival probability of 0.16. These results significantly surpass previously achieved performance in similar platforms. We further demonstrate the integration of these protocols with quantum error correction, with a logical error rates of $8.1(2)\times 10^{-3}$ per QEC round. 

The framework presented here, compatible with standard high-performance quantum error correction, addresses both state preparation and measurement errors, thus overcoming a practical challenge in building bosonic-based fault-tolerant quantum processors.

\section{\label{sec:methods}Methods}
\subsection{Experimental setup}
To experimentally demonstrate SPAM improvement and autonomous quantum error correction gain with single-mode grid state qubits, a superconducting device based on a cQED architecture is used. As shown in Fig. 1(a), the bosonic mode is hosted in a double-post cavity fabricated from high-purity aluminum, allowing for multi-mode operations. Here, only a single mode is used to encode information, hereafter referred to as the \textit{storage mode}. This mode has a single-photon relaxation time of \storagelifetimesymb \ = \storagelifetime\ and a coherence time \storagecoherencesymb \ = \storagecoherencetime. A superconducting chip containing a transmon auxiliary qubit, a readout resonator, and a Purcell filter is inserted into the microwave cavity. This provides the nonlinearity required for control operations and readout of the storage mode. The transmon qubit has a relaxation time of \transmonlifetimesymb = \transmonlifetime \ and an echo \transmonechosymb \ = \transmonecho. Due to mutual participation in the Josephson junction, the auxiliary qubit and the storage mode are dispersively coupled with \dispersiveshiftsymb \ = \dispersiveshift. This configuration enables high-fidelity preparation and manipulation of logical states encoded in the storage mode, described in the next section. For additional information on the measurements of these parameters, refer to Appendix \ref{apx: charact device}. 

\begin{figure}[t]
  \includegraphics[width=1\linewidth]{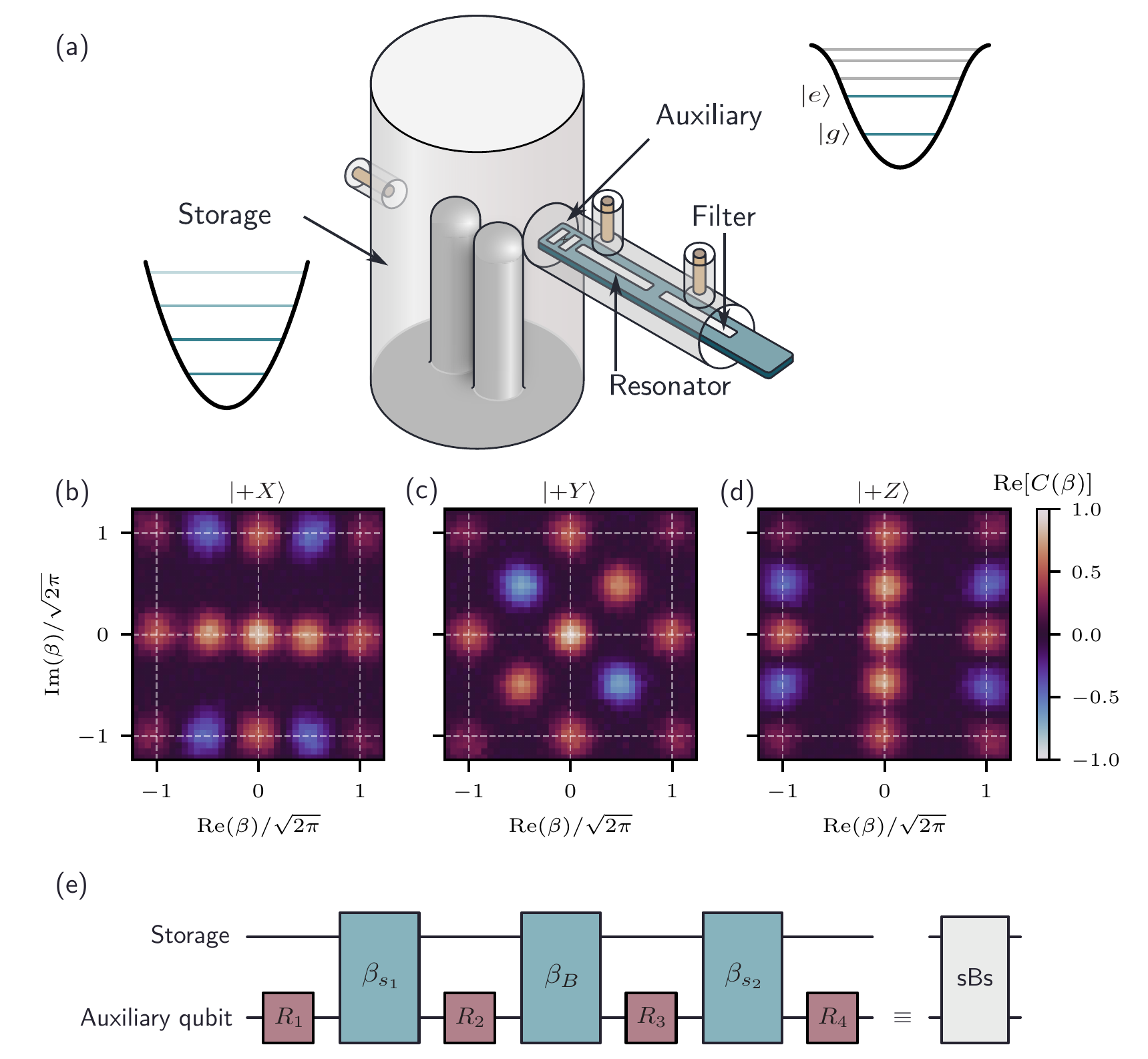}
  \caption{\textbf{Experimental system.} (a) Schematic of experimental device. The single-mode grid code is encoded in one of the fundamental modes of a double-post microwave cavity. The storage is dispersively coupled to an auxiliary transmon qubit. The auxiliary qubit is also dispersively coupled to an on-chip resonator and Purcell filter used for reset and readout.
  (b)-(d) Measured characteristic function ($\mathrm{Re}[C(\beta)]$) tomography of three cardinal states with $\Delta = 0.38$ through a postselected stabilization protocol starting from the storage mode in vacuum. (e) sBs stabilization conditional displacements and auxiliary qubit rotations used for quantum error correction.}
  \label{fig:fig1}
\end{figure}
\subsection{Grid-state encoding}
The logical information is encoded in the storage mode in a finite-energy single-mode grid state. The state is initialized with a finite-energy envelope, which shapes the amplitude distribution of the grid state, determining both the width of the individual peaks and the overall Gaussian decay. We use an envelope parameter of $\Delta = 0.38$, corresponding to an average photon number of $\simeq 2.96$, given by $\bar{n}_\mathrm{s} \approx (1-\Delta^2)/2\Delta^2 $. The size of the envelope also controls the resilience of the encoded grid state to errors and the rate at which they can be corrected through stabilization. 

\subsubsection{Quantum error correction}

\begin{figure*}
  \includegraphics[width=\textwidth]{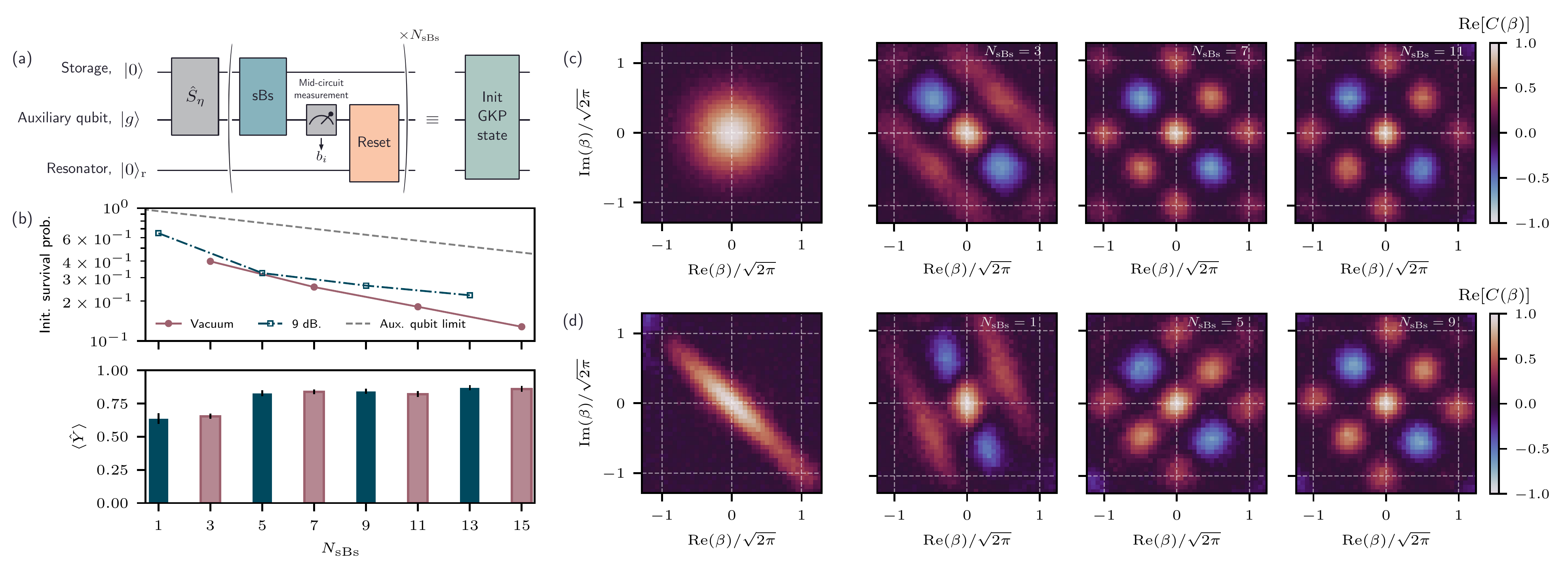}
  \caption{\textbf{State preparation from postselected stabilization.} (a) State preparation protocol from postselected stabilization. Each round corresponds to sBs stabilization unitary followed by a mid-circuit measurement and an auxiliary qubit reset. $\hat{S_\eta}$ refers to the squeezing operation performed using an optimized sequence of ECD and auxiliary qubit rotations.
  Mid-circuit measurement outcomes $\{b_i\}$ are used to perform postselection based on a all-agree to $\ket{g}$ policy. (b)
  Starting from vacuum (pink) or 9 dB squeezed (blue) state, experimental results for preparation of $\ket{+Y}$ with $\Delta = 0.38$ versus the number of stabilization rounds $N_\mathrm{sBs}$. Top panel shows measured survival probability with all auxiliary outcomes $b_i = g$ and the limit on this probability set by auxiliary readout fidelity of $\mathcal{V} = 0.95(1)$ (dashed). Bottom panel shows measured Pauli expectation value $\langle\hat{Y}\rangle$ via a single round of finite-energy measurement.
  (c)-(d) Characteristic function tomography at increasing numbers of rounds $N_\mathrm{sBs}$ from (b) after starting in a vacuum or squeezed state respectively.}
  \label{fig:fig2}
\end{figure*}
Errors such as photon loss can affect the logical grid state encoded in the storage mode. To protect and correct the logical states, we employ the sBs stabilization protocol \cite{royer_stabilization_2020}, which implements an effective finite-energy dynamics using operations experimentally accessible in our architecture. Each sBs round consists of four auxiliary qubit rotations $\mathbf{R}_\mathbf{sBs} = (+\frac{i\pi}{2}, +\frac{\pi}{2}, -\frac{\pi}{2}, -\frac{i\pi}{2})$, interleaved by echoed conditional displacements \cite{eickbusch_fast_2022, lachance-quirion_autonomous_2024}, parametrized by ($\beta_{s_1}, \beta_B, \beta_{s_2}$), followed by a reset of the auxiliary qubit (see Fig. \ref{fig:fig1}e). 

The stabilization drives the storage mode toward the GKP code-state manifold, enabling both error correction of an encoded logical state and state preparation from vacuum or a squeezed state. By tuning the size of the small and large displacements, we can control the grid-state envelope ($\Delta_\mathbf{sBs}$) and set the average number of photons of the encoded state with stabilization, at round $n_\mathrm{r}$, with 
\begin{align}\label{eqs:sBs disp}
    \beta_{s_{1}}^{(n_\mathrm{r})} &= \frac{i^{(n_\mathrm{r}-1)}\ell_{n_\mathrm{r}} \sinh{\Delta_\mathbf{sBs}^2} }{2\sqrt{2}}, \\  \beta_B^{(n_\mathrm{r})}  &= \frac{i^{n_\mathrm{r}}\ell_{n_\mathrm{r}} \cosh{\Delta_\mathbf{sBs}^2} }{\sqrt{2}},
\end{align}
where $\ell_{n_\mathrm{r}}$ is the length of the displacement operator. The second small displacement $\beta_{s_{2}}^{(n_\mathrm{r})}$ follows the same expression as Eq.~\ref{eqs:sBs disp}, up to a scaling factor such that  $\beta_{s_{2}}^{(n_\mathrm{r})}  = \zeta_{s} \beta_{s_{1}}^{(n_\mathrm{r})}$. In standard QEC of square grid states, the displacement operators corresponds to stabilizers with lattice length $\ell_{n_\mathrm{r}} = 2\sqrt{\pi}$. 

\subsubsection{Autonomous implementation via auxiliary reset}

Quantum error correction through stabilization can be performed fully autonomously without the need for feedback operations \cite{lachance-quirion_autonomous_2024}. This can be  achieved by resetting the auxiliary qubit via driving the $|f0\rangle \leftrightarrow |g1\rangle$ transition preceded by an auxiliary qubit $\pi_{ef}$ rotation. This effectively performs a dissipative swap between the transmon and the readout resonator. In practice, the reset sequence is composed of two swaps to effectively reset the auxiliary qubit states $\ket{e}$ and $\ket{f}$, with an averaged reset error $\langle \epsilon_\mathrm{reset}\rangle = 0.013(4)$ in \resettotalduration\ (See section~\ref{apx:reset}). 

\subsection{Postselected state preparation}
Typically, single mode grid state preparation relies on an optimized sequence of ECD and auxiliary qubit rotations. This method, successfully used in recent experiments \cite{eickbusch_fast_2022, sivak_real-time_2023, lachance-quirion_autonomous_2024}, is susceptible to auxiliary qubit errors during ECD. 

Here, we implement a protocol where state preparation is performed using stabilization, starting from either the vacuum or a squeezed state of the storage mode. This protocol offers precise control over the state's finite-energy envelope and target initial cardinal state. Each stabilization round is interleaved with a mid-circuit measurement and reset of the auxiliary qubit as illustrated in Fig. \ref{fig:fig2}(a). The measurement outcomes are used to implement a procedure compatible with a repeat-until-success approach, in which postselection keeps only instances where the auxiliary qubit is measured in its ground state after each round. Starting from a squeezed state, which is an eigenstate of one of the code stabilizers, can improve the state-preparation survival probability.

The mid-circuit measurement sequence includes a readout, a wait time of \preechodelayInit \ to empty the resonator, an auxiliary qubit $\pi$ pulse and an additional wait time of \postechodelayInit \ implementing a readout echo on the storage mode, followed by a reset of the auxiliary qubit. The readout echo is applied to cancel the storage-mode rotation induced by auxiliary qubit-state-dependent frequency shifts, ensuring consistent evolution across successive measurement rounds. The reset ensures robustness to auxiliary qubit readout errors and removes the need for feedback operations. 

\subsubsection{\label{sec:prep card grid}Preparation of cardinal grid-states}
We prepare cardinal states of the single-mode grid code by repeatedly applying stabilization cycles on a storage initial state. One stabilization cycle consists of several stabilization rounds, and the initial state may be vacuum or a squeezed state. The stabilization cycles are different for different cardinal states.

For states $|\pm X\rangle$ and $|\pm Z\rangle$, each quantum error correction cycle consists of three sBs rounds. Each cycle is composed of two stabilizer rounds and one Pauli round where big displacement in the QEC round correspond to either a stabilizer or Pauli operator respectively. The small displacements are scaled proportionally  to the large displacement (see Eq.~\ref{eqs:sBs disp}). The inclusion of a Pauli round in the stabilization cycle requires a gauge update to ensure proper stabilization of the encoded state. This requirement arises because the displacements enacted by the ECD in the Pauli round anti-commute with the orthogonal stabilizer~\cite{royer_encoding_2022}. The gauge update at every round is implemented \textit{via} a sign change in the final auxiliary rotation and the second small displacement ($\beta_{s_2}$). Details of the gauge update are provided in Appendix~\ref{apx:gauge update}. 

The $|\pm Y\rangle$ states are prepared by exploiting their equivalence to the rotated qunaught state ($|\text{\O}\rangle$). The qunaught state, which encodes no logical information, is the simplest square-grid state with lattice spacing $\ell_{\text{\O}} = \sqrt{\pi/2}$. In practice, we have $|\pm Y\rangle = e^{\pm i\frac{\pi}{4}\hat a^\dagger \hat a}\hat{T}(s_{x,\text{\O}}/2)|\text{\O}\rangle$, where $\hat{T}(s_{x,\text{\O}}/2)$ is a translation of half a stabilizer of the qunaught state in the $\hat{x}$ direction and $\hat a ^{(\dagger)}$ refers to the annihilation (creation) operator of the storage mode. These states are initialized using a cycle of two stabilizer sBs rounds ($\{\hat{S}_{x, \text{\O}}, \hat{S}_{p, \text{\O}}\}$), with the large displacements corresponding to the stabilizers of a rotated qunaught state at $\pm 45^\circ$ angle for $|\pm Y\rangle$ respectively. The stabilizers of the rotated qunaught state are equivalent to the two different representatives of the Pauli $\hat{Y}$ operators, $\{\hat{S}_{x, \text{\O}}e^{i\frac{\pi}{4}\hat{a}^\dagger \hat{a}}, \hat{S}_{p, \text{\O}}e^{i\frac{\pi}{4}\hat{a}^\dagger \hat{a}}\}\rightarrow \{\hat{Y}_{1}, \hat{Y}_{2}\}$, establishing a direct parallel with the inclusion of Pauli rounds in the preparation of $\ket{\pm X}$ and $\ket{\pm Z}$. For the preparation of $\ket{\pm Y}$ states, the gauge update must also be applied during stabilization (Appendix~\ref{apx:gauge update}).

\subsubsection{Presqueezing}
For both of the strategies described above, we compare two different initial conditions on the storage mode: a vacuum state and a squeezed state. When starting from a squeezed state, an optimized ECD sequence is applied to prepare a 9 dB–squeezed state in the storage mode. Presqueezing reduces the number of sBs rounds required to achieve high-fidelity grid states as the initial state is already an eigenstate of one of the code stabilizers and the corresponding Pauli operator. This approach thus improves the survival probability during postselection for a given target preparation fidelity.

\subsubsection{Example of $\ket{+Y}$ state preparation}
We first quantify the impact of state preparation errors and the performance of the sBs protocol by initializing the six cardinal grid-states. Here, we highlight the case of state $\ket{+Y}$, and compare the results from postselected stabilization with the standard state preparation procedure consisting of an optimized sequence of auxiliary qubit rotation and ECD gates \cite{eickbusch_fast_2022, sivak_real-time_2023, lachance-quirion_autonomous_2024}.

As shown in Fig.~\ref{fig:fig2}(c)-(d) for state $|+Y\rangle$, the state rapidly converges towards the grid-state manifold, when performing sBs stabilization, steering both vacuum and squeezed initial states toward the target state. In both cases, the Pauli expectation values increase and then saturate within a few rounds. The overall fidelity here is primarily limited by the auxiliary-qubit readout fidelity $\mathcal{V} =0.95(1)$ (see Appendix \ref{apx: mcm calibration}). For the results presented in this figure, Pauli operators values are measured using a single finite-energy measurement, and mid-circuit measurement outcomes are postselected on error-free state preparation. Pauli logical measurements are performed at specific number of stabilization rounds to ensure that the end state is in the trivial gauge (see Appendix~\ref{apx:gauge update}). While comparable Pauli expectation values are obtained for vacuum and squeezed initial states, the latter exhibits a higher survival probability, as it is initially closer to the code space, being an eigenstate of one of the stabilizers. To measurably discern the fidelity difference between preparation protocols, as a single-round of finite-energy measurement used here is not sufficient, we must improve logical Pauli measurements. This is addresses in section \ref{sec:Logical meas}, where a improved protocol with repeated measurements are implemented.

\begin{figure}[ht!]
  \includegraphics[width=0.95\linewidth]{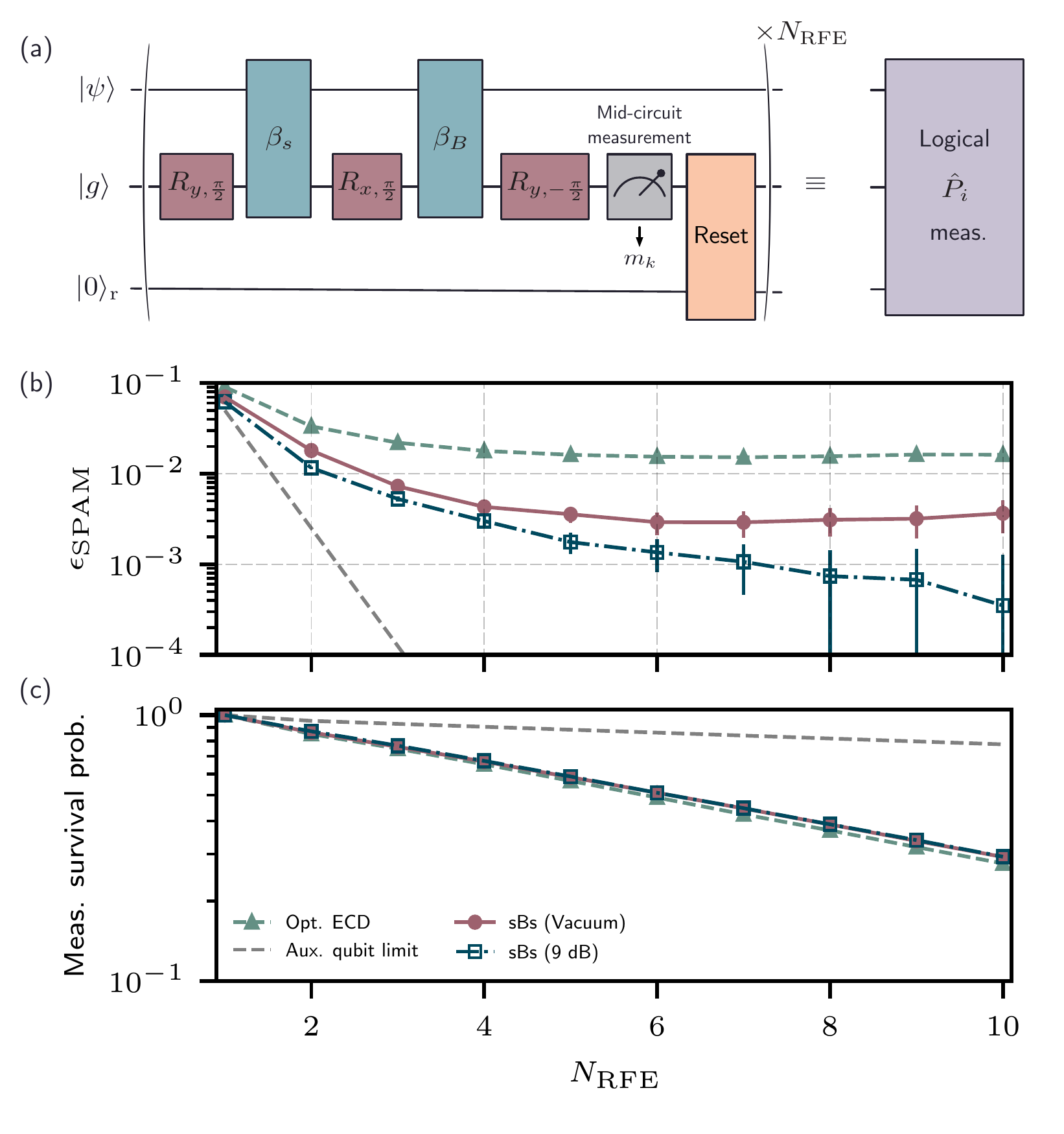}
  \caption{\textbf{Repeated finite-energy measurement.}
  (a) Repeated finite-energy measurement protocol that implements a single-shot logical Pauli measurement of $\hat{P_i}$. The ECD amplitude $\beta_B$ selects the desired $\hat{P_i} \in \{ \hat{X}, \hat{Y}, \hat{Z} \}$, and the finite-energy parameter is set to $\Delta = 0.38$. To maximize fidelity, an all-agree policy on the set of $N_\mathrm{RFE}$ measurements $\{m_k\}$ is applied and results in a sub-unity shot-survival probability.
  (b) Total state preparation and measurement error ($\epsilon_\mathrm{SPAM} = 1-F_\mathrm{L}$) averaged over all six prepared cardinal states as function of the number of tomography rounds $N_\mathrm{RFE}$ for three different state preparation protocols. Here, \textit{Opt. ECD} corresponds to an optimized ECD with a depth of 7. (c) Corresponding survival probability for data in (b). The dashed gray line corresponds to the limit set by the auxiliary qubit readout error on SPAM error and survival probability respectively.}
  \label{fig:fig3}
\end{figure}
\begin{figure*}[ht!]
  \includegraphics[width=\textwidth]{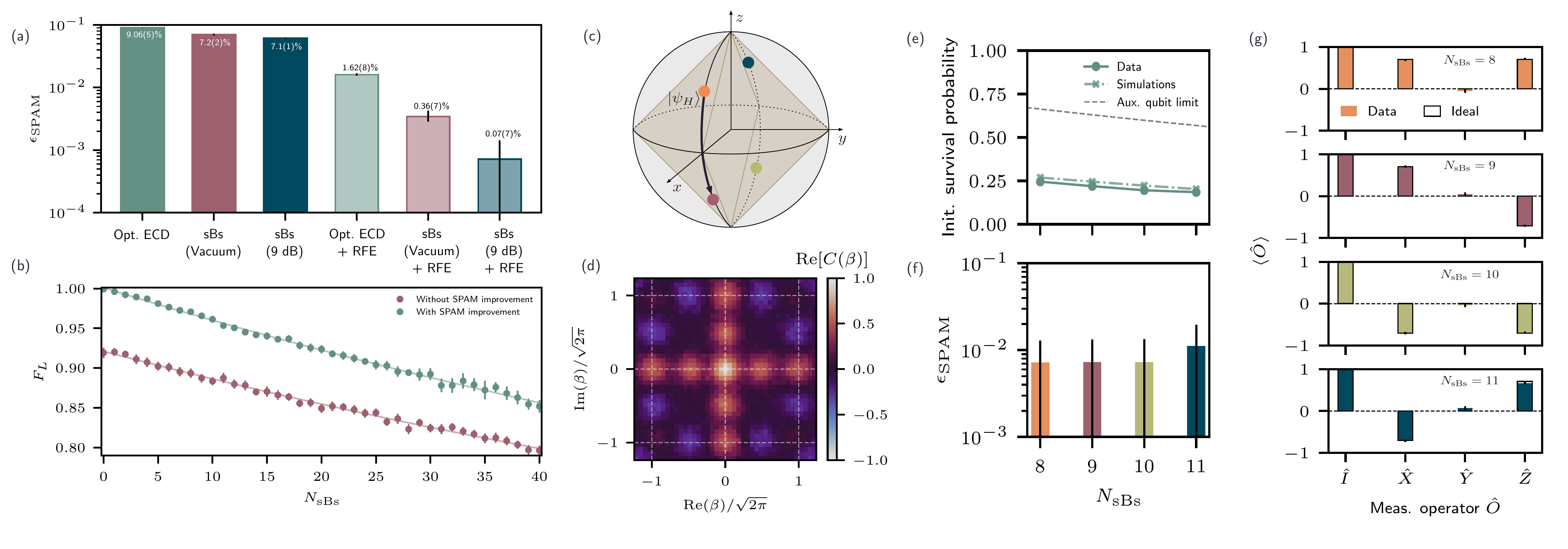}
  \caption{\textbf{SPAM improvement, QEC performances and magic state preparation.} (a) Total state preparation and measurement error ($\epsilon_\mathrm{SPAM} = 1-F_\mathrm{L}$) as function of the different combination of state preparation and measurement strategies.  For state preparation from stabilization protocol, the total SPAM error is obtained after 5 (8) rounds of repeated finite-energy (RFE) measurements starting from vacuum (9 dB squeezing). Postselection is performed based on an all-agree policy. (b) Evolution of the single-mode grid state.  The lines show fits used to extract the logical error per round under stabilization, yielding values of $0.0081(2)$ (green) and $0.0085(2)$ (pink) for QEC with and without SPAM improvement protocols in the restless regime, respectively. Without SPAM improvement refers to the initialization based on optimized sequences of ECD and a single round of finite-energy. (c) Bloch sphere representing the four $H$-type magic states initialized. (d)  Characteristic function tomography of magic state $\ket{\psi_H} = \frac{\ket{+Z} + \ket{+X}}{\sqrt{2 + \sqrt{2}}}$ (orange dot), with $\Delta = 0.38$. The state preparation corresponds to 8 rounds of postselected stabilization. (e)-(f) Survival probability and total SPAM error as function of the number of stabilization rounds. Each stabilization round does a $R_{\hat{Y}, \pi/2}$ rotation on the Bloch sphere thus preparing a different H-state. (g) Expectation values of operators $\hat{O}$ for the four different magic states initialized, measured with 5 rounds of finite-energy measurements.}
  \label{fig:fig4}
\end{figure*}
\subsection{Logical measurement}\label{sec:Logical meas}

Pauli expectation values are measured using the finite-energy measurement protocol shown in Fig.~\ref{fig:fig3}(a), which limits errors arising from the finite-energy envelope of the grid state compared to a simpler one-bit phase-estimation protocol \cite{royer_stabilization_2020, hastrup_improved_2021, singh_towards_2025}. We extend this approach by performing repeated finite-energy measurements within a single shot, combined with an all-agree postselection strategy, in which outcomes are retained only when all repetitions yield the same mid-circuit measurement result. From these postselected outcomes, the auxiliary qubit excited-state probability is reconstructed and mapped onto the logical Pauli expectation value. 

In practice, only a small number of finite-energy measurement rounds are required to saturate the Pauli expectation values, thereby preserving a high survival probability (Fig. \ref{fig:fig3}b). Both simulations and experimental data indicate that measurements of $\hat{X}$ and $\hat{Z}$ operators exhibit slightly higher survival probabilities than $\hat{Y}$, which can be attributed to their shorter displacement lengths in phase space. To symmetrize measurement errors, successive rounds of $\hat{X}$ or $\hat{Z}$ measurements are performed with a storage phase offset of $\pi$, while for $\hat{Y}$, both representatives are measured, corresponding to an effective phase offset of $\pi/2$.

\subsubsection{SPAM error metric}
To quantify the impact of SPAM errors experimentally, the logical fidelity is measured through the Pauli expectation values of six cardinal states of the single-mode grid code. The logical fidelity corresponds to 
\begin{equation}
    F_\mathrm{L} = \frac{1}{2} + \frac{1}{12}\sum_{\hat{\mu}_0 \in \{\hat{X}_0, \hat{Y}_0, \hat{Z}_0\}} (\langle \hat{\mu}_0 \rangle_+ - \langle \hat{\mu}_0 \rangle_-),
\end{equation}
where $\langle\hat{\mu}_0\rangle_\pm = \langle\pm \mu|\hat{\mu}_0|\pm \mu\rangle$ denotes the Pauli expectation value of state $|\pm \mu  \rangle$, with $ \mu \in \{X, Y, Z \}$. From this definition, the total SPAM errors is expressed as $\epsilon_\mathrm{SPAM} = 1-F_\mathrm{L}$, where the logical fidelity is measured for a given combination of state preparation and measurement protocols.

\subsubsection{Results}
As the number of repeated finite-energy measurement rounds increases, the total SPAM error decreases and rapidly saturate after only a few repetitions, as shown in Fig.~\ref{fig:fig3}(b). 

For a single round, SPAM fidelity is primarily limited by readout errors of the auxiliary qubit, which are significantly suppressed through repetition and postselection. As shown by the dashed grey line in Fig.~\ref{fig:fig3}(b), the experimental results of SPAM errors are no longer limited by the auxiliary qubit readout fidelity when two or more rounds of finite-energy measurements are performed. 

Figure~\ref{fig:fig3}(c) shows the survival probability following state preparation for different initialization methods, including an optimized ECD sequence and postselected stabilization-based state preparation from vacuum and squeezed state. The measurement survival probability is relatively independent of the initialization procedure and lower than the limit set by the auxiliary qubit readout fidelity alone. In Appendix~\ref{apx:RFE_exp}, we investigate the dependence of the survival probability on different postselection strategies. The analysis shows a trade-off between survival rate and SPAM performance and indicates strong correlations between measurement errors, which make the standard majority-vote strategy sub-optimal.

\subsection{Total SPAM performances}

Figure~\ref{fig:fig4}(a) provides a summary of the SPAM improvement protocols investigated in this work. As discussed above, results without repeated finite-energy are primarly limited by the auxiliary qubit readout fidelity. Nonetheless, a slight increase of performance is observed for stabilization-based state preparation. When combining this state preparation protocol with repeated finite-energy measurements, performance is further improved. The results of state preparation from postselected stabilization, starting from a squeezed state, and eight rounds of finite-energy measurement show a large reduction of the total SPAM errors reaching $7(7)\times 10^{-4}$. For state preparation and measurement using all-agree policies, the most restrictive postselection strategy, the total survival probability is 9\%, with a state preparation survival probability of 24\% and 39\% for measurement. This approach enables a substantial reduction of SPAM errors with minimal overhead, establishing repeated finite-energy measurements as a  practical tool for high-fidelity readout of grid states. 

\subsection{Quantum error correction}
The state-preparation and measurement improvement protocols introduced in this work, namely stabilization-based initialization and repeated finite-energy measurements, are compatible with high-performance autonomous quantum error correction. To demonstrate this, we implement error correction using stabilization-based sBs cycles in the restless regime, i.e., without idle time between consecutive rounds. Each sBs rounds has a duration of \timeofsBsround. 

The logical fidelity $F_\mathrm{L}(n_\mathrm{r})$ as function of the number of stabilization rounds is extracted from measurements of the Pauli expectation values of the six cardinal states. The decay of the logical fidelity is described by an exponential model,
\begin{equation}
    F_\mathrm{L}(n_\mathrm{r}) = \frac{1}{2} + \left(F_\mathrm{L}(0) - \frac{1}{2}\right) e^{-\epsilon_\mathrm{QEC} n_\mathrm{r}},
\end{equation}
where $F_\mathrm{L}(0)$ denotes the initial logical fidelity after state preparation and $\epsilon_\mathrm{QEC}$ is the logical error per stabilization round.

We focus on the short-time dynamics of the stabilization process, which is the most relevant for quantum computation. We compare two cases, as exposed in Fig.~\ref{fig:fig4}(b), both in which the stabilization is operating in a fully autonomous regime without mid-circuit measurements or postselection. In the first case, the system is initialized using an optimized ECD sequence and measured using a single finite-energy measurement. This yields a logical error per round of $\epsilon_\mathrm{QEC} = 0.0085(2)$. In the second case, the SPAM-improvement protocols are incorporated prior to autonomous error correction, resulting in $\epsilon_\mathrm{QEC}^\mathrm{SPAM\ imp.} = 0.0081 (2)$.

The two cases exhibit nearly identical logical error rates per round, demonstrating that the SPAM-improvement protocols do not degrade the performance of autonomous quantum error correction. In contrast, a significant enhancement of the initial logical fidelity $F_\mathrm{L}(0)$, enabled by postselected stabilization and repeated finite-energy measurements, reflects the possible substantial improvements in both state preparation and measurement. 

These results demonstrate that SPAM-improvement protocols introduced above can substantially enhance the quality of initial states and measurements without compromising error-correction performance, providing a clear pathway toward high-fidelity operations.

\subsection{Magic state preparation}
The ability to generate high-fidelity magic states is a key asset for universal quantum computation, enabling the execution of non-Clifford gates without directly implementing non-Gaussian operations \cite{gupta_encoding_2024,daguerre_experimental_2025}. In this work, we show that such states can be prepared starting from the vacuum with only minimal modifications to the stabilization-based initialization protocol introduced above. The vacuum state, serves here a simple and readily available resource for magic states state preparation \cite{baragiola_all-gaussian_2019}.

To prepare $H$-type magic states, we apply repeated quantum error-correction cycles based on the two stabilizers of the square grid-state qubit ${S_x, S_z}$. The stabilization is applied to a storage mode initially in vacuum, postselected on no-error states. Each round of error correction effectively implements a logical $\hat{X}$ or $\hat{Z}$ operation, thereby cycling the encoded state through the set 
$\ket{\psi_H} \propto \ket{\pm Z} \pm \ket{\pm X}$, as illustrated in Fig.~\ref{fig:fig4}(c). Additionally, in contrast to teleported-gate protocols \cite{campagne-ibarcq_quantum_2020}, this protocol does not require feedback operations.  

Figures 4(e)–(f) display the state preparation survival probability and total SPAM error for the four H-type magic states prepared using this protocol. The SPAM error is defined as $\epsilon_\mathrm{SPAM} = 1- F_{|\psi\rangle}$, where the fidelity is extracted from a reconstructed state obtained via the measurement of the following operators $\left(\hat{I}, \hat{X}, \hat{Y}, \hat{Z}\right)$ expectation values. The state $\ket{\psi_H} = \frac{1}{\sqrt{2 + \sqrt{2}}}(\ket{+Z} + \ket{+X})$, prepared after eight rounds, achieves the best performance with $\epsilon_\mathrm{SPAM}= 8(5)\times 10^{-3}$. The associated uncertainty is estimated using a bootstrap resampling method, with a 95\% confidence interval.

Magic states prepared with number of rounds higher than eight exhibit higher SPAM errors and reduced survival probabilities, consistent with the accumulation of imperfections over successive stabilization rounds. Figure 4(g) shows the corresponding expectation values for each state. 

These results highlight how the vacuum state provides a particularly effective resource, when combined with our SPAM improvement protocols, for high-fidelity magic-state preparation.

\section{Conclusion}
In conclusion, we demonstrated high-fidelity state preparation and measurement of grid states in a superconducting cavity coupled to an auxiliary transmon qubit. By combining postselected stabilization-based state preparation and repeated finite-energy measurements, we achieved total SPAM errors below $10^{-3}$. These protocols are fully compatible with autonomous quantum error correction and do not deteriorate its performance. Furthermore, the vacuum state serves as a simple yet powerful resource for preparing high-fidelity H-type magic states with total SPAM errors of $8\times 10^{-3}$. To push performances further, these strategies can be extended to multi-mode codes offering higher protection against errors and a scalable route towards fault-tolerant quantum computation\cite{royer_encoding_2022, lemonde_hardware-efficient_2024}. 

\begin{acknowledgments}
S.T. and B.R. acknowledge support from NSERC, the Fonds de recherche du Québec – Nature
et technologie and the Canada First Research Excellence Fund.
\end{acknowledgments}
\setcounter{figure}{0}
\setcounter{table}{0}
\renewcommand{\thefigure}{\thesection.\arabic{figure}}
\renewcommand{\thetable}{\thesection.\arabic{table}}
\appendix
\section{Experimental method} \label{apx: charact device}
\subsection{Experimental device}

The experimental device is composed of a superconducting double-post microwave cavity machined in high-purity Al. A superconducting chip made on a sapphire substrate with a strip line resonator, a Purcell filter and a transmon is coupled to the storage mode through the waveguide leading to the storage cavity. The resonator, filter and transmon are made of tantalum except for aluminum junction. The superconducting chip is held inside the waveguide and thermally anchored with a copper clamp. The coupling between the transmon auxiliary qubit and the storage modes is controlled by the insertion of this chip inside the waveguide. 

For more information about the device assembly procedure, we refer the reader to Ref.~\cite{lachance-quirion_autonomous_2024}.

\begin{table*}[t]
\centering
\begin{tabular}{c|l|c|l}
\hline\hline
\textbf{Mode} & \textbf{Parameter} & \textbf{Value} & \textbf{Method of estimate or measurement} \\
\hline
\multirow{6}{*}{\shortstack{Storage\\mode}}
 & Dispersive shift \dispersiveshiftsymb & \dispersiveshift & Out-and-back experiment \cite{eickbusch_fast_2022} \\ 
 & Second-order disp. shift \seconddispshiftsymb & \seconddispshif & Out-and-back experiment \\ 
 & Kerr non-linearity \kerrsymb & \kerr & Out-and-back experiment \\ 
 & Relaxation time \storagelifetimesymb & \storagelifetime & Fock state $\ket{1}$ lifetime measurement\\ 
 & Coherence time \storagecoherencesymb & \storagecoherencetime & Fock state $\ket{0}+\ket{1}$ state coherence measurement \\
\hline
\multirow{6}{*}{\shortstack{Auxiliary\\mode}}
 & Anharmonicity \qubitanharmsymb & \qubitanharm & Two-tone spectroscopy of $\ket{g}\rightarrow\ket{e}$ and $\ket{e}\rightarrow\ket{f}$ transition\\ 
 & Thermal population $n_\mathrm{th}^\mathrm{q}$& 0.005(4) & Equilibrium population of the auxiliary qubit measurement \cite{lachance-quirion_autonomous_2024}\\
 & Relaxation time \transmonlifetimesymb & \transmonlifetime & Relaxation time measurement \\
 & Ramsey coherence time \transmonramseysymb & \transmonramsey &  Ramsey measurement\\
 & Echo coherence time \transmonechosymb & \transmonecho & Echo measurement \\
\hline
\multirow{2}{*}{\shortstack{Readout mode}}
 & Dispersive shift \qubitresonatordispshiftsymb & \qubitresonatordispshift & Direct RF reflection measurement\\
 & External coupling \kapparsymb & \kappar & Direct RF reflection measurement\\
 \hline
\multirow{1}{*}{\shortstack{Filter mode}}
 & External coupling \kappafsymb & \kappaf & Direct RF reflection measurement\\
\hline
\end{tabular}
\caption{\textbf{Summary of device parameters.}}\label{tab:parameters}
\end{table*}

\subsection{Hamiltonian parameters}
In the operating regime, where each modes are far detuned from each-other, the static part of the system can be described by the following dispersive Hamiltonian,
\begin{align*}
    \hat{H}_\mathrm{disp.}/\hbar&  \ = \ \underbrace{ \omega_\mathrm{s} \hat{a}^\dagger \hat{a} +  \omega_\mathrm{q} \hat{b}^\dagger \hat{b} + \omega_\mathrm{r} \hat{c}^\dagger \hat{c}}_{\text{dressed \ modes}} \\
    & + \underbrace{\frac{K_\mathrm{s}}{2}\hat{a}^{\dagger 2} {\hat{a}}^2 + \frac{K_\mathrm{q}}{2}\hat{b}^{\dagger 2} {\hat{b}}^2 + \frac{K_\mathrm{r}}{2}\hat{c}^{\dagger 2} {\hat{c}}^2}_{\text{dressed self-Kerr}}  \\
    & + \underbrace{2\chi_\mathrm{sq} \hat{a}^\dagger\hat{a} \hat{b}^\dagger \hat{b} + 2\chi_\mathrm{qr} \hat{b}^\dagger\hat{b} \hat{c}^\dagger \hat{c}}_{\text{cross-Kerr}}  \\
    & + \underbrace{2 \chi_\mathrm{s^2 q} \hat{a}^{\dagger 2}\hat{a}^2 \hat{b}^\dagger \hat{b} + 2 \chi_\mathrm{sq^2} \hat{a}^{\dagger }\hat{a} \hat{b}^{\dagger 2} \hat{b}^2 + 2 \chi_\mathrm{qr^2} \hat{b}^{\dagger}\hat{b} \hat{c}^{\dagger 2} \hat{c}^2}_{\text{second order cross-Kerr}}.
\end{align*}
Here, $\hat{a}$, $\hat{b}$, and $\hat{c}$ denote the annihilation operators for the dressed storage mode, auxiliary qubit, and readout resonator, respectively, with corresponding dressed angular frequencies $\omega_\mathrm{s}$, $\omega_\mathrm{q}$, and $\omega_\mathrm{r}$. The auxiliary qubit is modeled as a Kerr oscillator characterized by a self-Kerr $K_\mathrm{q}$. It is dispersively coupled to both the storage and readout modes, with interactions described up to first- and second-order cross-Kerr terms, while the anharmonicities of the storage and readout modes are given by $K_\mathrm{s}$ and $K_\mathrm{r}$, respectively. The Purcell filter mode is not considered here. We refer the reader to Ref.~\cite{lachance-quirion_autonomous_2024} for more in depth description of the system parameters.

\subsection{Auxiliary qubit reset}\label{apx:reset}
\begin{figure}[ht!]
    \centering
    \includegraphics[width=0.95\linewidth]{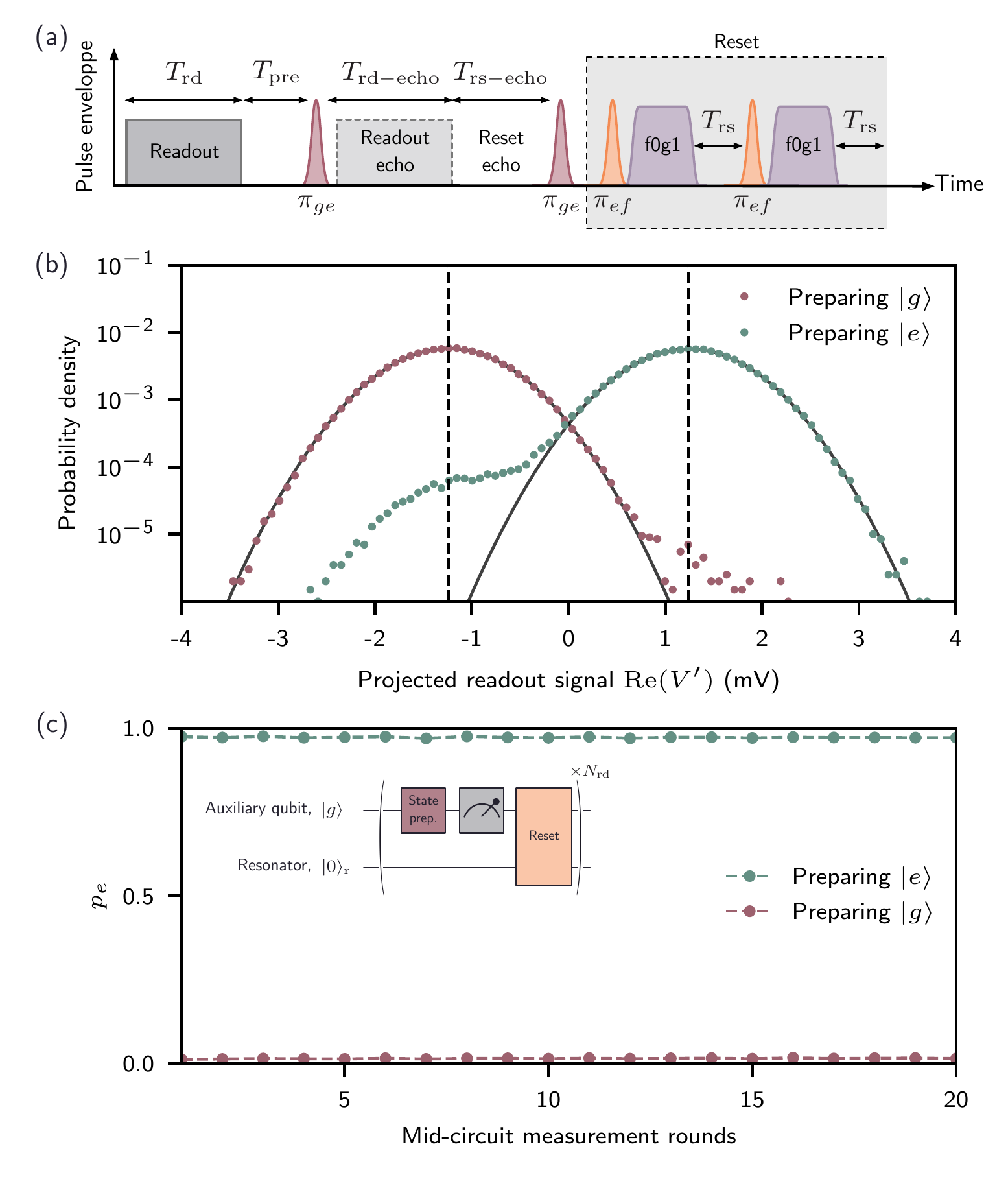}
    \caption{\textbf{Mid-circuit measurement calibration.} (a) Schematic of single round of mid-circuit measurement and reset protocol. The readout is followed by a pre-echo delay ($T_\mathrm{pre}$), a $\pi_{ge}$ pulse, a readout echo delay ($T_\mathrm{rd}$), a reset echo delay ($T_\mathrm{rs-echo}$), a second $\pi_{ge}$ pulse and an auxiliary qubit reset (see Table \ref{tab:mcm timings}).  (b)  Histogram of the projected readout signal $\mathrm{Re}(V^\prime)$ for $N=10^5$ shots when preparing the auxiliary ground and excited states $\ket{g}$ (pink) and $\ket{e}$ (green), respectively. Solid lines are Gaussian fits to show the readout signals corresponding to $\mathrm{Re}(V_g^\prime)$ and $\mathrm{Re}(V_e^\prime)$ (vertical dashed lines). (c) Auxiliary qubit excited state probability when preparing $\ket{g}$ (pink) and $\ket{e}$ (green) as function of the number of mid-circuit readout and reset pairs. Inset shows the protocol used for the measurements of the auxiliary qubit as function of the number of mid-circuits measurement rounds ($N_\mathrm{rd}$).}
    \label{fig:A.4 mcm calib}
\end{figure}

As mentioned in the main text, the auxiliary qubit reset is performed through a drive between the $|f0\rangle \leftrightarrow |g1\rangle$ of the qubit-resonator sub-system. The complete reset sequence is composed of two identical blocks of pulses. Each block is composed of a $\pi_{ef}$ pulse on the auxiliary qubit, followed by a microwave \textit{f0g1} drive between the $|f0\rangle \leftrightarrow \ket{g1}$ transition and a reset delay to empty the resonator. The sequence is illustrated in Fig.~\ref{fig:A.4 mcm calib}(a). 

Experimentally, the $\pi_{ef}$ operation is performed with a Gaussian pulse of full width at half maximum of 10 ns, the \textit{f0g1} drive has a width of 20 ns and a plateau of 55 ns and the reset delay is of 295 ns. This results in a total reset time of 800 ns. 

The reset error is defined by $\epsilon_\mathrm{reset} = \frac{1}{3}\sum_{k\in\{g,e,f\}} (1-p_{g|k})$, where $p_{g|k}$ represent the probability of measuring the auxiliary qubit in the ground state after the reset, given that it was initially prepared in state $\ket{k}$. 

The presence of photons in the storage mode can impact the reset performance. To capture this effect, the reset error is measured and averaged over three coherent states in the storage mode, corresponding to average photon numbers of $\bar{n} \in \{0, 2, 4 \}$. The reset parameters, such as the \textit{f0g1} drive amplitude, frequency, drive duration, and reset delay, are then optimized to minimize the average reset error across these three coherent states. The averaged reset error obtained experimentally is $\bar\epsilon _\mathrm{reset} = 1.3(4) \times 10^{-2}$.

\subsection{Mid-circuit measurements calibration}\label{apx: mcm calibration}
\begin{table}[ht!]
\begin{tabular}{c|c|c}
\hline \hline
 & \shortstack{State preparation\\from postselected\\stabilization} & \shortstack{Repeated\\finite-energy\\measurements}  \\
 \hline
Readout ($T_\mathrm{rd}$) & 1 $\mu$s & 1 $\mu$s \\[1ex]
\makecell{ Readout pre-echo \\ delay ($T_\mathrm{pre}$)} & 220 ns & 220 ns \\[10pt]
$\pi_{ge}$ width  & 10 ns & 10 ns \\[1ex]
\makecell{Readout echo \\ delay ($T_\mathrm{rs-echo}$)} & 180 ns & 1.18 $\mu$s \\
\hline
\end{tabular}
\caption{\textbf{Summary of mid-circuit measurement timings per SPAM improvement protocols.}}\label{tab:mcm timings}
\end{table}
The readout of the auxiliary is benchmarked by evaluating the readout visibility, defined as 
\begin{equation}
    \mathcal{V} = p_{g|g} - p_{g|e},
\end{equation}
where $p_{g|k}$ is the probability of assigning the readout result to $\ket{g}$ when preparing in a state $\ket{k}$ with $k\in {g, e}$.

The mid-circuit measurements of the auxiliary qubit are performed using the protocol illustrated in Fig~\ref{fig:A.4 mcm calib}(a). Following the resonator readout, the sequence consists of a readout pre-echo delay, a $\pi_{ge}$ pulse, a readout echo delay, a reset echo delay, a second $\pi_{ge}$ pulse, and a final reset sequence. The pre-echo delay allows the resonator to empty after the readout drive, while the interval between the two $\pi_{ge}$ pulses compensates for the storage rotation induced by the auxiliary qubit state-dependent frequency. The timing parameters used in each protocol are summarized in Table~\ref{tab:mcm timings}. 

The readout fidelity in the context of the mid-circuit measurements is calibrated using sequences of repeated auxiliary state preparation, readout, and reset operations. The measurements shown in Fig.~\ref{fig:A.4 mcm calib} demonstrate that the reset procedure does not degrade the auxiliary qubit readout fidelity, which remains constant over repeated cycles. This indicates that, after measurement in a given state, the qubit is effectively reset and yields consistent outcomes upon subsequent state preparation and readout. From these measurements, we extract a mid-circuit readout fidelity of 0.95(1).

\subsection{ECD gate}
In the experiments, the echoed conditional displacement (ECD) gates are implemented over a total duration of \ecdtotaltime, comprising displacement pulses of duration \displacementwidth, interaction time \ecdinteractiontime, and $\pi$ pulses of duration \pipulsewidth. The amplitude of the final displacement pulse in the ECD sequence is scaled by a factor $\zeta_\mathrm{ECD}$, calibrated to cancel any residual unconditional displacement. This scaling depends on the ECD timing and device-specific parameters, and is optimized experimentally. 

The ECD drive frequency is calibrated using an \textit{out-and-back} experiment \cite{eickbusch_fast_2022}. From this experiment, parameters such as the dispersive shift between the storage mode and the auxiliary qubit are extracted as described in Table~\ref{tab:parameters}. In addition, the phase accumulation of the qubit during the ECD operation is characterized using a cat-and-back protocol \cite{sivak_real-time_2023}. The calibration of the ECD amplitude, relating the drive amplitude to the resulting displacement in phase space, is performed via measurements of the vacuum characteristic function. Further details on these calibration procedures can be found in Refs.~\cite{lachance-quirion_autonomous_2024, eickbusch_fast_2022, sivak_real-time_2023}. In practice, the ECD parameters are refined through a closed-loop optimization performed for each protocol, following the procedure described in Ref.~\cite{lachance-quirion_autonomous_2024}.

To assess the overall ECD gate performance, the auxiliary qubit is initialized in the ground state $\ket{g}$, after which an ECD operation is applied and the induced displacement is subsequently undone using an unconditional displacement. This sequence is repeated multiple times, and the resulting signal is fitted to an exponential decay as a function of the number of iterations. From this, we extract an ECD gate error of $\epsilon_{\mathrm{ECD}} =$ \ecdfidelity.

\subsection{Storage presqueezing state preparation}
Presqueezing of 9 dB is implemented using a numerically optimized sequence of ECD gates and auxiliary qubit rotations. A circuit with depth $N=7$, corresponding to the number of ECD and rotation pairs, is used here. The phase space orientation of the presqueezing is adjusted for the different target states. For $|\pm X\rangle$ states, the squeezing is oriented along the axis $\mathrm{Re}(\beta)$; for $|\pm Z\rangle$ states, along the axis $\mathrm{Re}(\beta)$; and for $|\pm Y\rangle$ states, along $\pm 45^\circ$ in phase space, respectively. 

\subsection{Characterization of the storage coherence times}
\begin{figure}[t]
    \centering
    \includegraphics[width=1\linewidth]{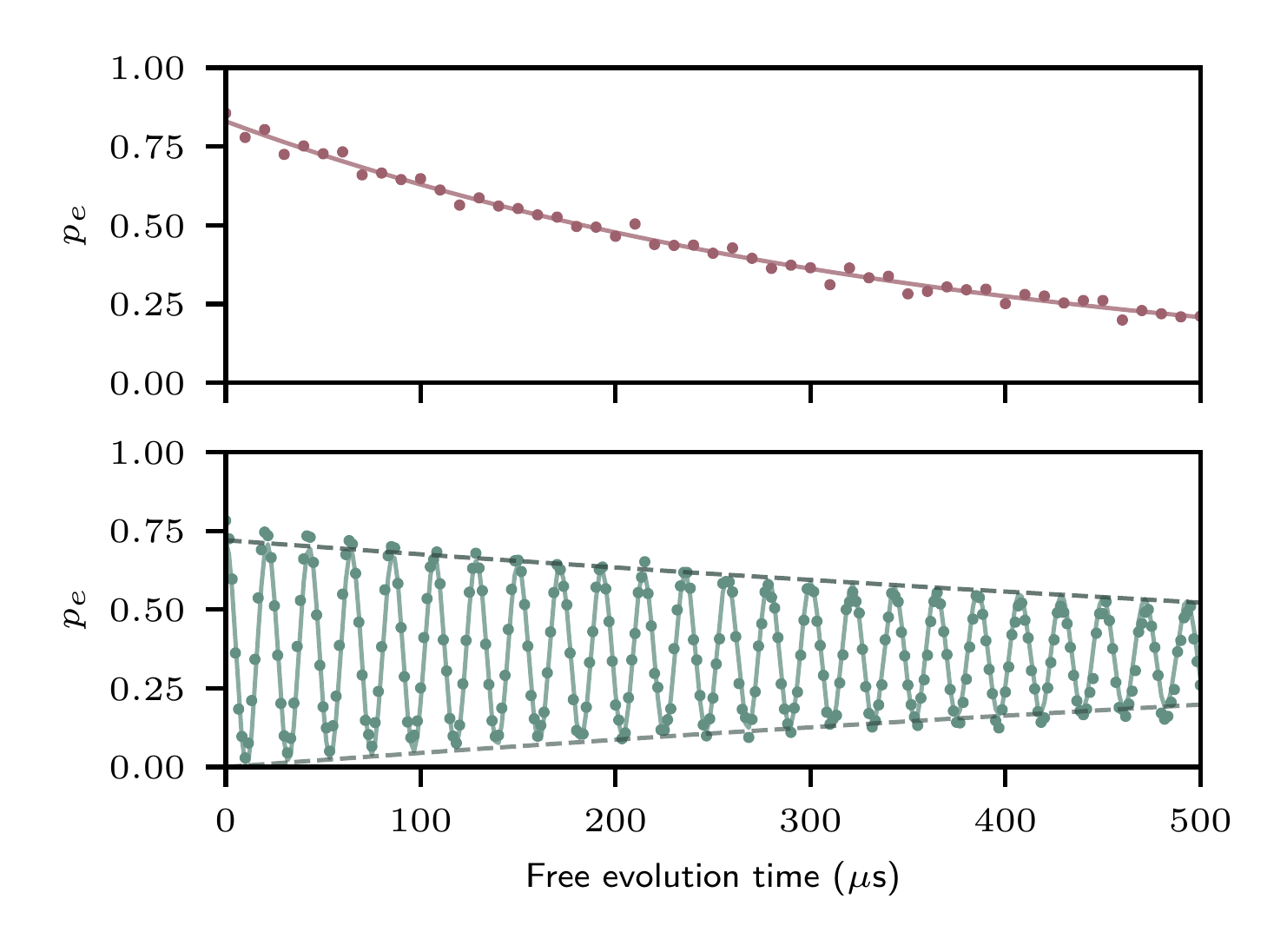}
    \caption{\textbf{Storage mode coherence times.} Relaxation time \storagelifetimesymb\ and \storagecoherencesymb\ of the $\{\ket{0}_\mathrm{s}, \ket{1}_\mathrm{s}\}$ Fock state qubit including SPAM errors. The coherence times are measured through the auxiliary qubit with a decoding sequence mapping $\ket{0}\rightarrow\ket{g}$ and $\ket{1}\rightarrow\ket{e}$.}
    \label{fig:B.1 lifetimes}
\end{figure}

To measure the relaxation time of the storage mode, the system is first initialized in a single-photon Fock state. After a free evolution time, the population of the single-photon state is measured by mapping the on photon state $\ket{1}$ onto the excited state of an auxiliary qubit, $\ket{e}$. This mapping, as well as the preparation of $\ket{1}$, are implemented using an optimized ECD sequence. As showed in Fig. \ref{fig:B.1 lifetimes}, the single photon relaxation time is \storagelifetimesymb = \storagelifetime.

For the storage mode coherence time measurement, we perform a Ramsey-type experiment. To do so, two logical $\hat{Y}_{\pi/2}$ gates for the \{01\} encoding are applied to the storage mode, interleaved by a variable free evolution time. The state is then decoded using the same sequence employed for the relaxation time measurement. The $\hat{Y}_{\pi/2}$ operations are realized via optimized ECD sequences. From the measurements, we extract a coherence time of \storagecoherencesymb = \storagecoherencetime, as shown in Fig. \ref{fig:B.1 lifetimes}. 

\section{Theoretical framework}\label{apx: noise model}
The theoretical framework used to support the experimental results of the main text and appendix are based on simulations using the noise model described in this section. Results from these simulations are presented in appendices~\ref{apx: Init. from stab.} and~\ref{apx: RFE} containing additional discussions on the state preparation by postselected stabilization and the repeated finite-energy measurement.

The noise model in simulations includes the main decoherence noise channels affecting the state preparation and measurement performance. It includes storage decay as well as transmon decay and dephasing. These channels are active during ECD simulations and storage idling corresponding to the duration of auxiliary mid-circuit measurements and resets. Each half of the ECD gate, composed of a single conditional displacement ($\mathrm{CD}(\beta)$), is simulated using the following Hamiltonian, $\hat H_\mathrm{CD} = \frac{1}{2T_\mathrm{CD}} \left(i \beta \hat a^\dagger - i \beta^* \hat a\right) \hat\sigma_z$, for a duration $T_\mathrm{CD}$. The auxiliary qubit operations (reset and rotations) are taken to be ideal and instantaneous.

This noise model captures and describes the main types of errors that can occur during primitive operations, but abstracts pulse-level dynamics to simplify time-evolution simulations. This produces performance upper bounds as it does not capture all noise sources.
\section{State preparation by postselected stabilization}\label{apx: Init. from stab.}
\setcounter{figure}{0}
\setcounter{table}{0}
In this section, we provide additional theoretical background on the state preparation by postselected stabilization based on simulations using the noise model described in appendix~\ref{apx: noise model}. Results of operator expectation values are also computed using this noise model with simulations of single finite-energy measurement (for repeated finite-energy measurement simulations, see appendix~\ref{apx: RFE}). The state preparation of $\ket{\pm Z}$ and $\ket{\pm Y}$ states are shown in appendix~\ref{apx: init sim} and $\ket{\pm X}$ states are obtained from $\ket{\pm Z}$ states under a $90^\circ$ storage rotation.

\subsection{Postselected stabilization cycles}\label{apx:init cycles}
The cycles of stabilization rounds are defined in this section. Based on the ordering of the different types of sBs rounds, the total number of rounds is chosen such that after a certain numbers of cycles, the state is in the trivial gauge (see appendix~\ref{apx:gauge update}), i.e. the gauge where the grid state has $+1$ value for stabilizers and $\pm 1$ for the corresponding Pauli. To simplify the notation, we identify each type of round with an infinite-energy stabilizer or Pauli operator. One must keep in mind that the corresponding sBs round implements the stabilization of the finite-energy operator based on its displacement vector in phase space.

For preparing $|\pm Z\rangle$ states, the cycle of sBs rounds follows the order $\{\hat S_x, \hat S_p, \hat Z\}$, i.e. first applying the two sBs rounds corresponding to the square-GKP stabilizers followed by a $\hat Z$ round, where the indices of the stabilizers indicates their displacement direction in phase space and $\hat Z$ corresponds to half the displacement of $\hat S_p$ in the same direction. For $\ket{\pm X}$ states, the same cycle is used with each operator rotated by $90^\circ$, $\{\hat S_p, \hat S_x, \hat X\}$.

For preparing state $\ket{+Y}$, the cycle is adjusted based on whether the storage state is initially squeezed or in vacuum. When starting from the vacuum, the cycle starts with the sBs round corresponding to $\hat Y_2$ stabilizer, resulting in the following QEC cycle $\{\hat Y_2, \hat Y_1\}$, where these operators are defined in section~\ref{sec:prep card grid}. When starting from a squeezed state, the squeezing is oriented to obtain a well defined eigenvalue for finite-energy $\hat Y_2$ and the cycle starts with the orthogonal round $\{\hat Y_1, \hat Y_2\}$.
The state $\ket{-Y}$ is obtained by applying a $90^\circ$ rotation to the stabilization operators used for state $\ket{+Y}$.

\subsection{Simulations}\label{apx: init sim}

\begin{figure}[t]
    \centering
    \includegraphics[width=\linewidth]{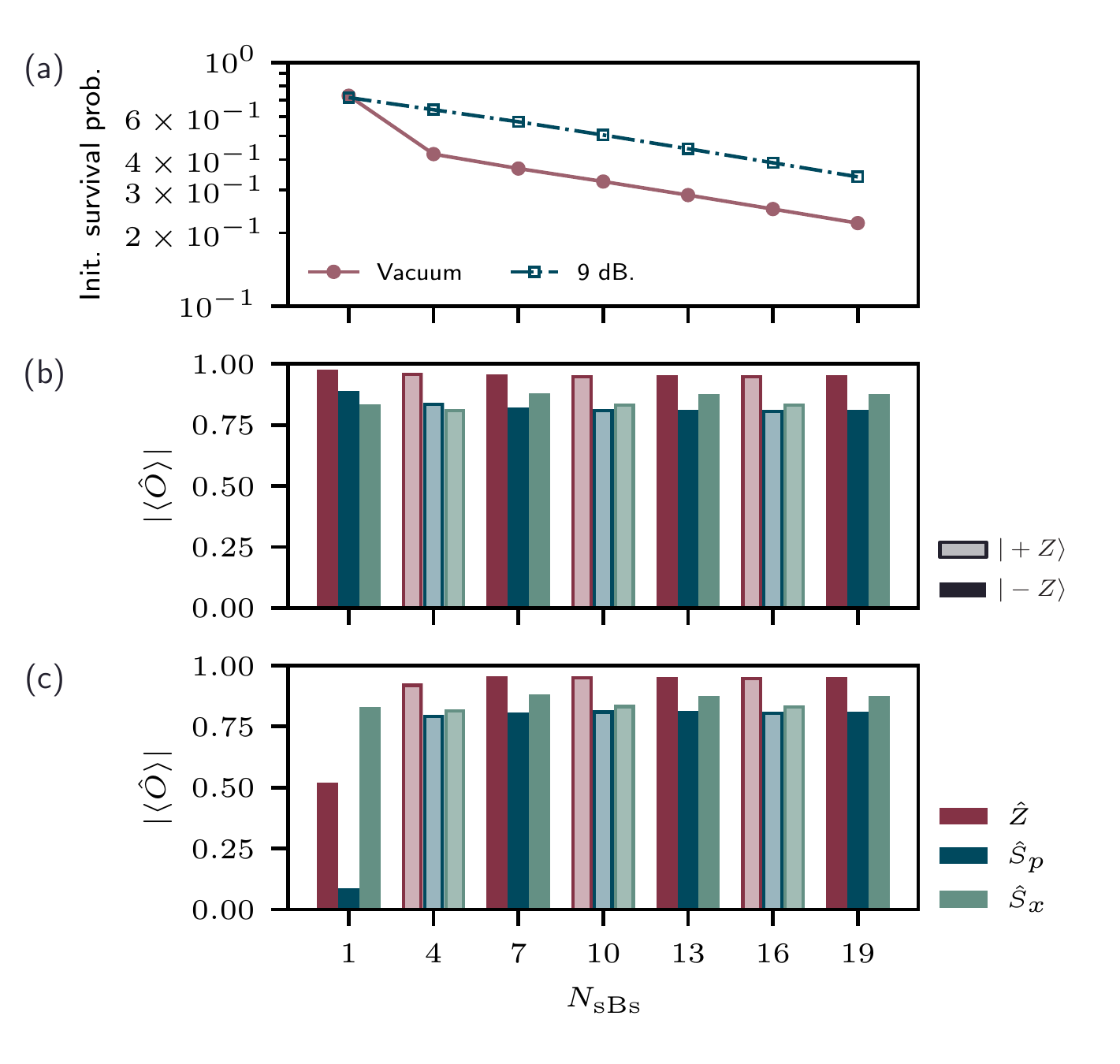}
  \caption{\textbf{Simulation results of state preparation from postselected stabilization of Z states.} (a) Survival probability as a function of the number of rounds starting from the vacuum (red) and from a squeezed state (blue), when post-selecting on the auxiliary measured in $|g\rangle$ after each round, assuming ideal readout fidelity. (b) Starting from a squeezed state and (c) starting from the vacuum, the absolute value of the expectation value in the measurement of the two finite-energy stabilizers and $\hat Z$, obtained in simulations of single finite-energy measurements. Darker (lighter) colors corresponds $|-Z\rangle$ ($|+Z\rangle$). The appended Pauli ($\hat Z$) round for $|+Z\rangle$ states (see section \ref{apx:gauge update}) is not included in these results where the main cycle is followed throughout the sequence.\label{fig:init Z combined}}
\end{figure}

In Fig.~\ref{fig:init Z combined}, simulation results of $\ket{\pm Z}$ state preparation starting from the vacuum and a squeezed state are presented. The survival probability (a) and the expectation value of $\hat Z$, $\hat S_p$ and $\hat S_x$ measurements (b)-(c) are shown as a function of the number of rounds. The squeezing allows to start the sequence with the logical information already encoded in eigenvalues of finite-energy $\hat Z$ and $\hat S_p$, making the survival probability of this approach higher than when starting from the vacuum. In both approaches, we observe $|-Z\rangle$ state (darker colors) reaching saturation in the three operators measurements at round 7. For $|+Z\rangle$ state (lighter colors), $\hat Z$ measurements reaches saturation at 10 rounds when starting from the vacuum, and at 4 rounds when starting from a squeezed state. This difference illustrates the second advantage of the initial squeezing, i.e. reducing the cost in the number of rounds which directly translates to an additional gain in survival probability.

\begin{figure}[ht!]
    \centering
    \includegraphics[width=\linewidth]{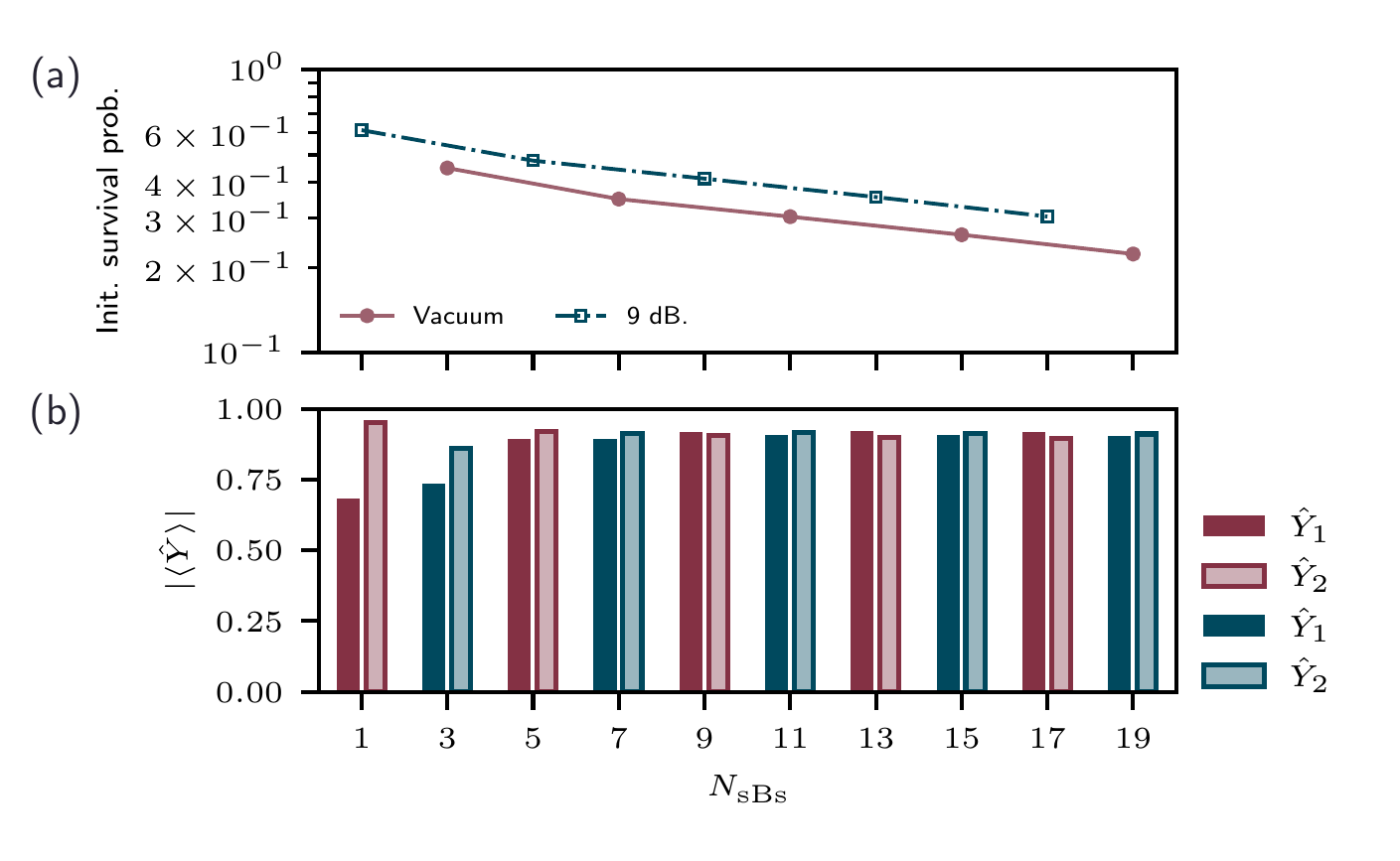}
  \caption{\textbf{Simulation results of state preparation from postselected stabilization of Y states from vacuum.} (a) Survival probability as a function of the number of rounds, when post-selecting on the auxiliary measured in $|g\rangle$ after each round, assuming ideal readout fidelity. (b) Absolute value of the expectation value in the measurement of the two finite-energy $\hat Y_1$ (darker color) and $\hat Y_2$ (lighter color). Both panels compare the approaches starting from a squeezed in red and the vacuum in blue. \label{fig:init Y combined}}
\end{figure}

Similar behaviors are observed with $\ket{\pm Y}$ states in Fig.~\ref{fig:init Y combined}, where the initial squeezing allows to start with a high expectation value of the finite-energy $\hat Y_2$. This leads to an increase in survival probability (a), with saturation for both $\hat Y_1$ and $\hat Y_2$ reached at round 9 and 11, when starting from squeezing and vacuum respectively. 

\subsection{Gauge}\label{apx:gauge update}

Our state preparation protocols contains sBs rounds where the big ECD corresponds to a stabilizer or a Pauli operator. Pauli rounds leave the state in a modified codespace, where the peaks are not on the original phase space lattice points, but are shifted by half a lattice spacing. We say that such a modified codespace has non-trivial gauge. We account for this gauge change in software by modifying subsequent sBs rounds. Depending on the precise relationship between the resulting gauge's vector and the vector of the subsequent sBs round~\cite{royer_encoding_2022}, the last displacement and last rotation of the subsequent round may need to be flipped.

In this work, we make the choice to end up with a grid state in the trivial gauge, which imposes constraints on the total number of rounds. Note that alternatively, it is possible to prepare the state in any gauge (e.g., for maximizing the logical fidelity) and keeping track of it in software with updates in the subsequent operations and measurements.

The QEC cycles during state preparation is implemented such that they respect the constraint stated above while minimizing the total number of stabilization rounds required to prepare a given state. 

For $\ket{\pm X}$ and $\ket{\pm Z}$ states, Pauli rounds need to be applied an even number of times to ensure the $+1$ value for stabilizers.
The preparation of state $\ket{+Z}$ requires the first stabilizer of the QEC cycle ($\hat S_x$, which is orthogonal to the Pauli) to be applied an even number of times. To satisfy these two constraints, the full QEC cycle ($\{ \hat S_x, \hat  S_z, \hat Z\}$) is performed an odd number of times and followed by a single $\hat S_x$ and a $\hat Z$ round. This corresponds to a total number of rounds of $6n+5$, where $n$ is an integer.

For $\ket{-Z}$, the $S_x$ stabilizer needs to be applied an odd number of times. The full cycle is then performed an even number of time, followed by only a $S_x$ round. This corresponds to a total number of rounds of $6n+1$.

For $\ket{\pm X}$, the sequence follows the same structure by interchanging stabilizers $\hat S_x \leftrightarrow \hat S_z$ and replacing the Pauli round $\hat Z \rightarrow \hat X$.

For $\ket{\pm Y}$ states, the number of $\hat Y_1$ ($\hat Y_2$) rounds has to be odd (even), for a total number of rounds of $4n+3$ from the vacuum and $4n+1$ from a squeezed state, where the difference only comes from the different ordering of the two rounds between the two approaches.

\section{Repeated finite-energy measurement simulations}\label{apx: RFE}
\setcounter{figure}{0}
\setcounter{table}{0}

In this section, we investigate several strategies for improving the performance of repeated-measurements protocols. We begin with a detailed description of the enhanced $\hat{Y}$ measurement strategy, which relies on the use of two representatives of the Pauli $\hat Y$ operator. This exploration is carried out numerically using ideal finite-energy grid states. We then examine several post-selection strategies, considering both experimental data and numerical simulations. These results provide a means to tradeoff the total SPAM errors and the measurement survival probability.

\subsection{Improved $\hat{Y}$ measurements}
Here, we investigate how the choice of Pauli representative affects logical Pauli $\hat Y$ measurement in the repeated-measurement protocol. For Pauli $\hat Z$ and Pauli $\hat X$, the available representatives of each operator are collinear and therefore commute with a trivial zero symplectic phase. In contrast, Pauli $\hat Y$ admits two orthogonal representatives $\hat{Y}_1$ and $\hat{Y}_2$, see Sec.~\ref{sec:prep card grid},  that commute with a nonzero symplectic phase of $2\pi$.

Towards this goal, we define a two-outcome measurement observable:
$\hat{O}_Y = \hat{K}^{\dagger}_{+1,Y} \hat{K}_{+1,Y} - \hat{K}^{\dagger}_{-1,Y} \hat{K}_{-1,Y}$ where $\hat{K}_{+1,Y}$, $\hat{K}_{-1,Y}$ are Kraus operators associated with logical measurement outcomes $+1$ and $-1$, respectively, for a given Pauli $\hat{Y}$ representative. 
\begin{figure}[h!]
  \includegraphics[width=0.95\linewidth]{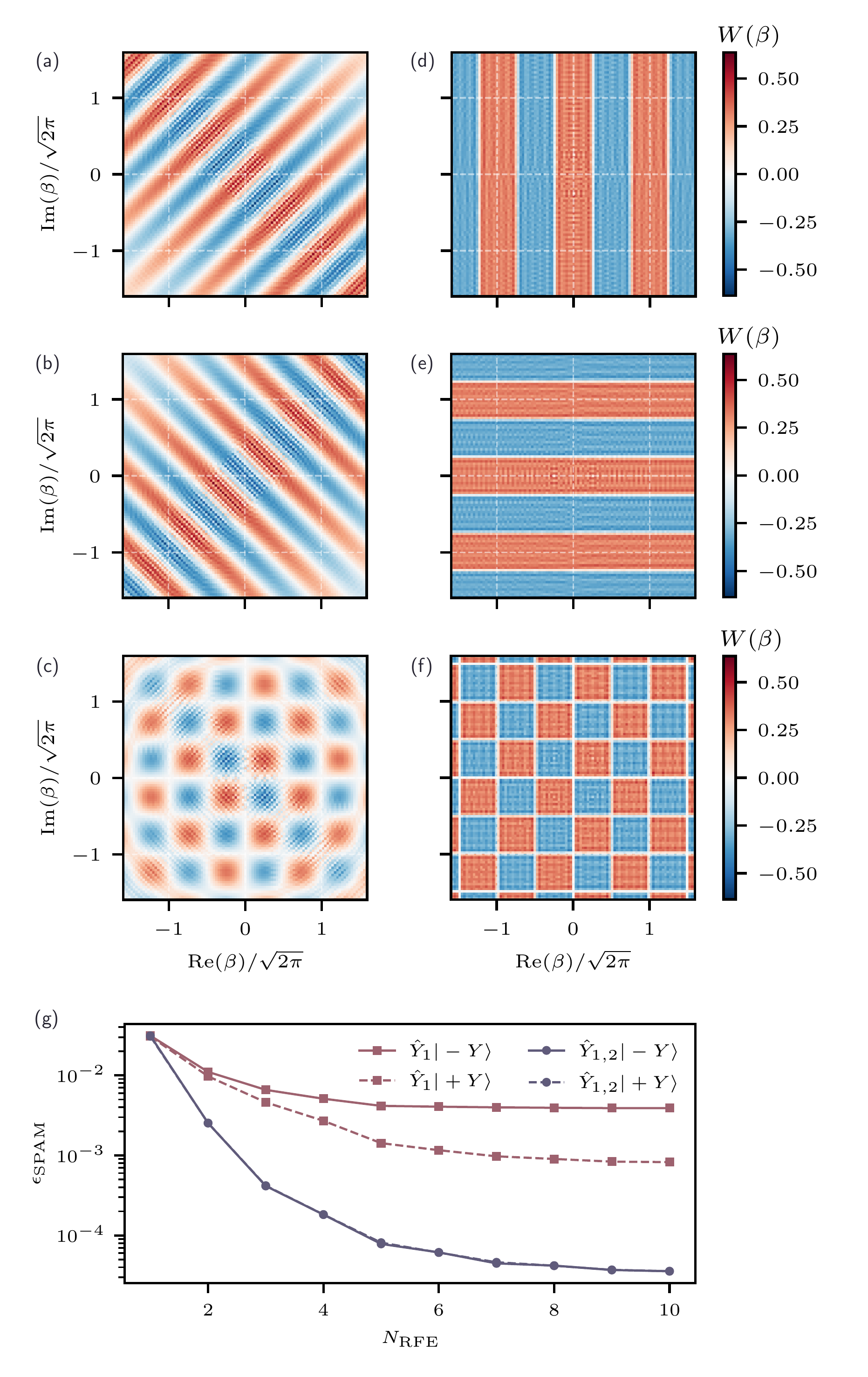}
  \caption{\textbf{Logical Pauli-$Y$ measurement using single and alternating representative measurement protocols.}  Wigner functions of the logical measurement observables for single-shot finite-energy measurements of the $\hat{Y}_1$ (a) and $\hat{Y}_2$ (b) respectively.
(c) Wigner function of the logical measurement observable $\hat{O}_Y$ corresponding to a single-shot finite-energy implementation based on consecutive measurements of $\hat{Y}_1$ and $\hat{Y}_2$. The outcomes $gg$ and $ee$ are mapped to the logical outcomes $+1$ and $-1$, respectively, while the mixed outcomes $ge$ and $eg$ are discarded through postselection.
(d)--(f) Wigner functions of the logical measurement observables for the numerically constructed measurement operators obtained from a truncated series expansion based on the subsystem decomposition for the $\hat{Z}$, $\hat{X}$, and $\hat{Y}$ Pauli operators, respectively.
(g) Measurement error ($\epsilon_\mathrm{SPAM}$) as function of the measurement round when one Pauli representative is measured per round.
The legend specifies the input logical $Y$ state and the representative protocol, either repeated $\hat{Y}_1$ measurements or alternating $\hat{Y}_1$ and $\hat{Y}_2$ measurements.
Simulations use the baseline noise model described in Sec.~\ref{apx: noise model} with $\Delta=0.38$ with ideal input states.}\label{fig:app_pauli_y_kraus_operators}
\end{figure}
The Wigner function of $\hat{O}_Y$ provides a phase-space representation of the measurement contrast: regions with positive and negative values are preferentially associated with the two logical outcomes. For an input state $\rho$, the expectation value $\langle \hat{O}_Y\rangle_{\rho}=\mathrm{Tr}(\hat{\rho} \hat{O}_Y)=p(+1|\hat{\rho})-p(-1|\hat{\rho})$ is equivalently determined by the phase-space overlap between the Wigner functions of $\hat{\rho}$ and $\hat{O}_Y$. Therefore, when the Wigner function of the measurement observable matches the checkerboard interference structure of the logical Pauli $\hat{Y}$ eigenstates, the two eigenstates produce a larger contrast in $\langle \hat{O}_Y\rangle_\rho$ and hence a lower logical measurement error probability. Conversely, missing or distorted phase-space features reduce this contrast and increase the probability of assigning the wrong logical outcome. Panels~(a)--(f) of Fig.~\ref{fig:app_pauli_y_kraus_operators} show the Wigner
functions of the experimental and theoretical measurement observables
considered here. 

In particular, panels~(d)--(f) correspond to the numerically constructed measurement operators obtained from a truncated series expansion based on the subsystem decomposition \cite{shaw_logical_2024, matsos_universal_2025}. This measurement can be interpreted as implementing a nearest-lattice-point decoder measurement. For logical $\hat{Z}$ ($\hat{X}$) states, this decoder yields a striped vertical (horizontal) measurement pattern, while for logical $\hat{Y}$ states, it yields a checkerboard structure. In each case, the pattern reflects the underlying structure of the corresponding logical states.

Panels~(a)--(c) correspond to the experimental protocol implemented using the finite-energy measurement circuit via an auxiliary transmon. In particular, panels~(a) and~(b) show single-shot measurements using the two Pauli $\hat{Y}$ representatives, $\hat{Y}_1$ and $\hat{Y}_2$, respectively. Although both operators are intended to measure the same logical Pauli $\hat{Y}$ operator, their phase-space structures are different. In particular, the corresponding measurement observables do not exhibit the full checkerboard pattern expected for the logical $Y$ eigenstates; instead, they display stripe-like features along opposite diagonals. Panel~(c) shows the measurement observable obtained by measuring both representatives and postselecting on consistent outcomes, with $gg$ and $ee$ assigned to logical outcomes $+1$ and $-1$, respectively, and $ge$ and $eg$ discarded. The resulting measurement observable recovers the symmetric checkerboard-like interference pattern, indicating that the combined protocol better resolves the logical Pauli $\hat{Y}$ states than either single representative alone.

This improvement is quantified in Fig.~\ref{fig:app_pauli_y_kraus_operators}(g). Repeated $\hat{Y}_1$ measurements reduce the SPAM error initially but quickly saturate and remain visibly dependent on the input logical $|\pm Y\rangle$ state. In contrast, the protocol relying on both $\hat{Y}_{1}$ and $\hat{Y}_{2}$ yields a substantially lower error, reaching below-$10^{-4}$ range after several rounds, while the two input logical eigenstates become nearly indistinguishable on the scale of the plot.

\subsection{Survival probability vs infidelity tradeoff}\label{apx:RFE_exp}
The postselection procedure becomes increasingly resource-intensive as system size grows due to the exponential decrease of the survival probability. For state preparation, this is not a problem since repeat-until-success approaches, which are fully compatible with the introduced protocols, can be implemented. Post-selection as part of measurement is on the other hand challenging in real algorithms as the survival probability decreases exponentially as function of the number of measurements. Hence, we study here the survival probability and fidelity trade-off in the context of the repeated finite-energy measurements as function of postselection policies.

\subsubsection{Postselection strategies}
Here, we investigate two alternative postselection policies that can increase the survival probability. The first and least restrictive policy is a majority-vote scheme. While this approach does not reject any measurement outcomes, the total SPAM errors are significantly higher than the other postselection strategies. This can be a result of the measurement back action due to the repeated measurements described in the following section (see Sec.~\ref{ref: RFE simulations tradeoff}). The second policy considered is a modified version of the all-agree strategy, in which a limited number of disagreeing outcomes are tolerated following a prescribed number of finite-energy measurements with agreeing outcomes. This postselection protocol is parameterized by the total number of measured bits $N_T$, the number of initial consecutive agreeing bits $N_{fa}$, and the maximum number of allowed errors $N_e$. As an example, for $N_{fa}=3$, $N_e = 1$ and $N_T = 5$, the postselection would keep : [$ggggg$, $gggeg$, $gggge$, $eeeeg$, $eeege$, $eeeee$].

\subsubsection{Experimental results}
\label{apx exp results}
We first look at the effect of three postselection strategies on the total SPAM errors and survival probability. This analysis is performed on the experimental data obtained from postselected initialization of cardinal states, starting from a squeezed state combined with repeated finite-energy measurements. In the main text, the most restrictive policy is applied, where all outcomes of repeated finite-energy must agree. This comes at the cost of reducing exponentially the measurement survival probability as the number of measurement increases. 

\begin{figure}[t!]
 \includegraphics[width=0.95\linewidth]{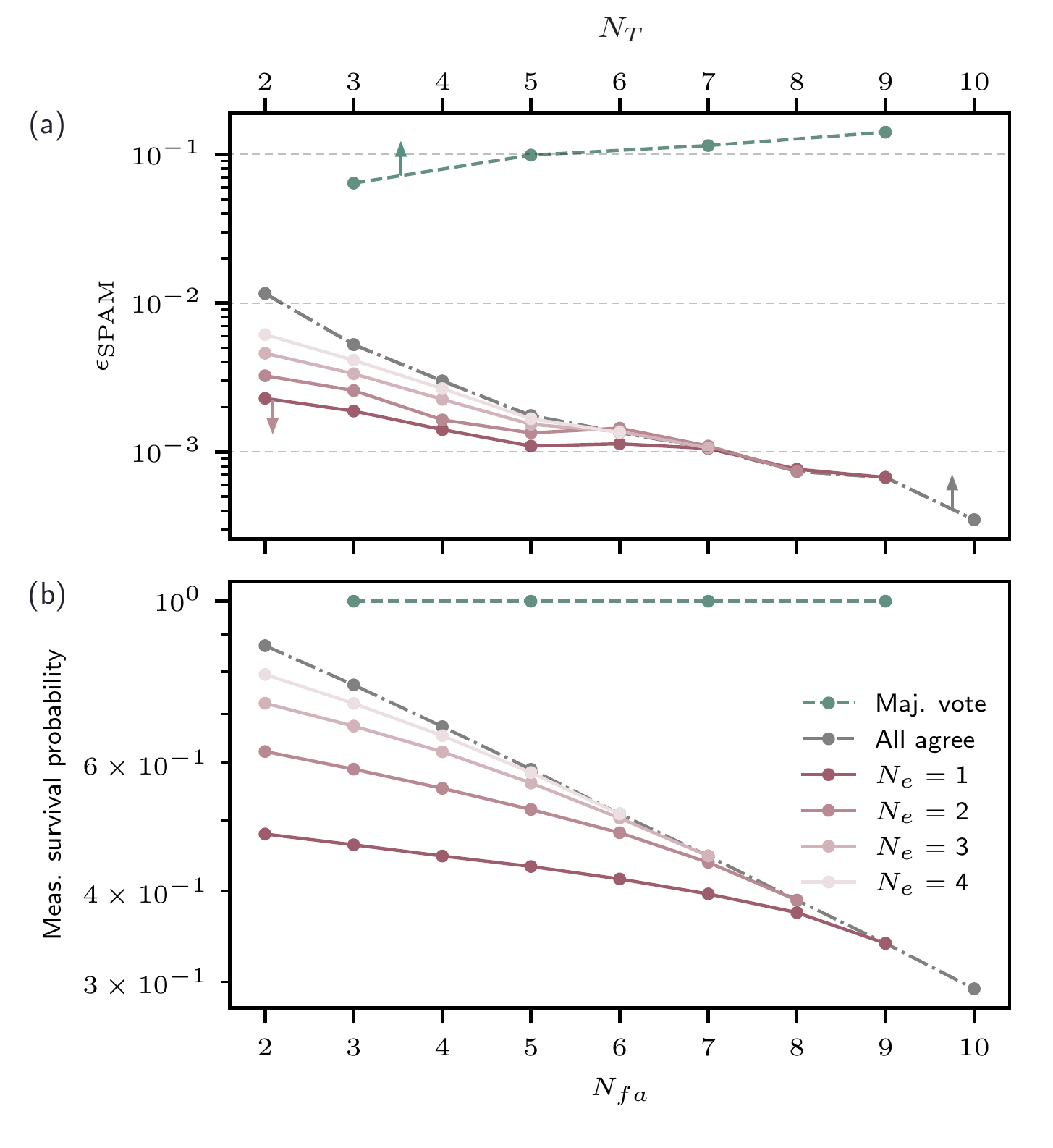}
  \caption{\textbf{Experimental postselection policy tradeoff analysis.} (a) Total SPAM error averaged over the six cardinal states as a function for different postselection strategies. The strategies considered are All-agree (grey), All-agree except $N_e$ (pink), and Majority vote (green). For the All-agree and Majority vote policies, the top x-axis represents the total number of measured bits. For the All-agree except $N_e$ strategy (pink), the x-axis denotes the number of first-agree bits ($N_{fa}$) out of a total of 10 rounds of repeated finite-energy measurements, where up to $N_e$	bit flips are allowed. Arrows denotes the x-axis of reference. (b) Measurement survival probability corresponding to the postselection strategies described above.}
  \label{fig:D2 exp. survival prob tradeoff - exp}
\end{figure}

Figure~\ref{fig:D2 exp. survival prob tradeoff - exp} illustrates the trade-off between the total SPAM error $\epsilon_\mathrm{SPAM}$ and the survival probability as a function of either the total number of measurements (for the all-agree and majority-vote policies) or the number of first agreeing bits (for the all-agree–except-$N_e$ policy). As discussed above, the majority-vote scheme yields a constant survival probability of unity, however, this comes at the cost of an increased $\epsilon_\mathrm{SPAM}$. The all-agree-except-$N_e$ policy is plotted for $N_e \in [1, 4]$ for a total number of bits $N_T = 10$. We compare the results to the all-agree policy as function of the total measured bits $N_T^{(aa)}$. 
In some cases, while the policy gives lower SPAM error, it also leads to a lower survival probability compared to the all-agree policy when $N_T^{aa} = N_{fa}$. However, there are some cases where the survival probability, when allowing few errors, surpass the all-agree policy while maintaining a total SPAM error below $10^{-2}$. In the case of the all-agree-except-$N_e$ policy, the survival probability also decays exponentially, but the decay rate is lower compared to the all-agree policy.

\subsubsection{Simulations}\label{ref: RFE simulations tradeoff}
\begin{figure}[th!]
    \includegraphics[width=0.95\linewidth]{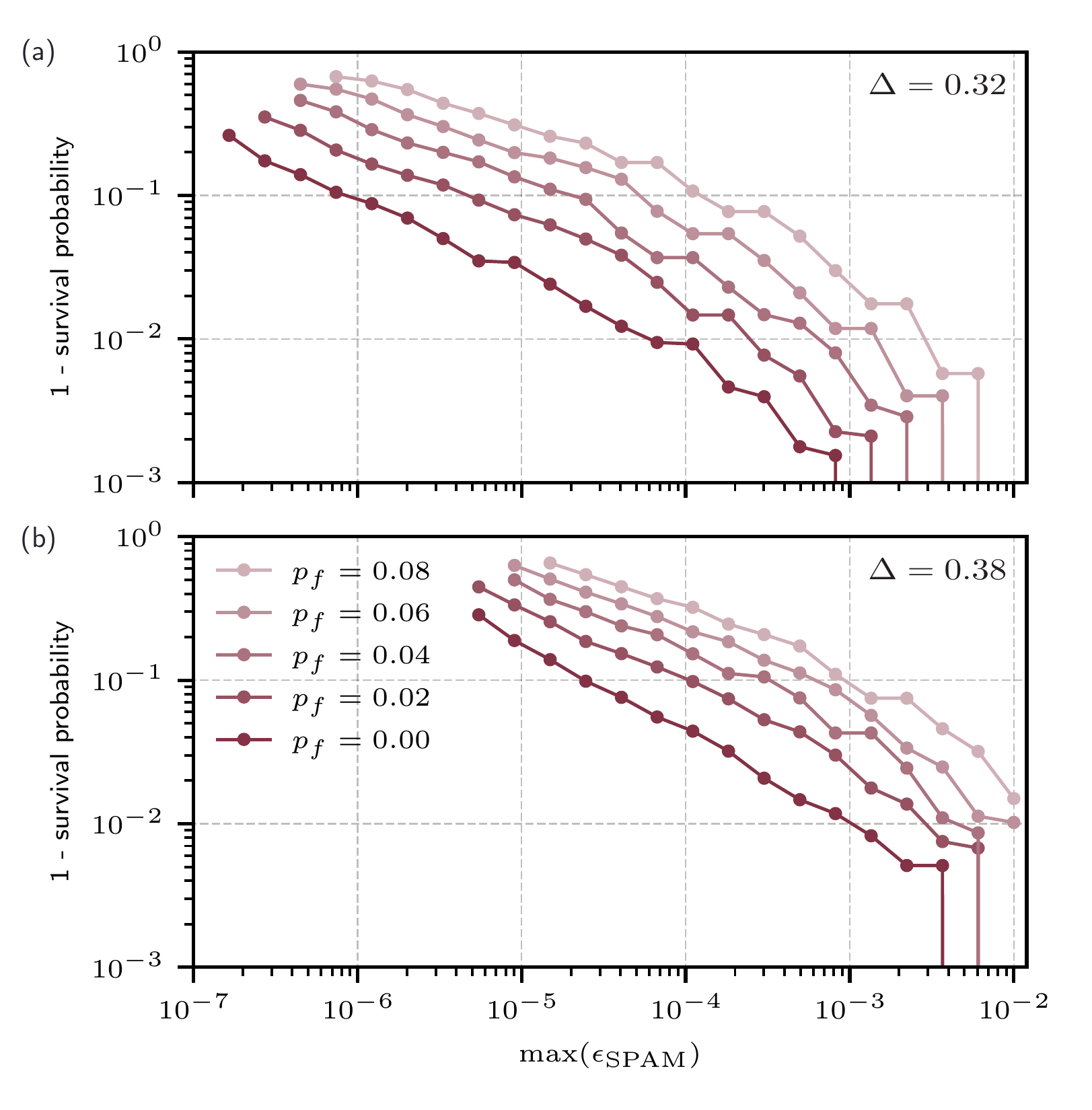}
  \caption{\textbf{Maximized measurement survival probability as a function of the allowed measurement error.} For each value on the x-axis, the measurement post-selection policy is optimized subject to the constraint that the measurement error does not exceed the specified threshold. The policy space is parameterized as described in Sec.~\ref{apx exp results}, with the total number of rounds constrained to $N_r \leq 10$. The top and bottom panels correspond to Pauli $\hat Z$ measurements of the ideal $\ket{0}$ state for $\Delta = 0.32$ and $\Delta = 0.38$, respectively. In addition to the baseline simulated noise model described in Sec.~\ref{apx: noise model}, we include a classical bit-flip channel on the final measurement outcome: with probability $p_f$, the binary measurement æresult is flipped, and with probability $1-p_f$, it is left unchanged. The value of $p_f$ is indicated in the legend.}
  \label{fig:app_survival_vs_targeted_infidelity}
\end{figure}

We now examine the trade-off between survival probability and measurement infidelity obtained from numerical simulations. The objective is to understand the influence of the finite-energy parameter $\Delta$ and the readout error on the survival probability. For this, we simulate 10 rounds of finite-energy measurements of the Pauli $\hat{Z}$ operator on the ideal $\ket{+Z}$ state. In particular, we compute the probability of all $2^{10}$ possible measurement bitstrings. Using these data, we maximize the survival probability for a prescribed level of logical measurement error ($\epsilon_\mathrm{SPAM}$) over the parameterized class of post-selection policies  described in the previous section. The resulting dependence of the optimal survival probability on the prescribed measurement error is shown in Fig.~\ref{fig:app_survival_vs_targeted_infidelity}.

We observe that our family of post-selection policies provides a controlled interpolation between the all-agree policy, which minimizes the measurement error, and the majority-vote policy, which maximizes the survival probability. By relaxing the requirement of achieving the lowest possible error, the survival probability can be increased systematically. At low measurement error, the dependence follows an approximately clean power law. Beyond a threshold, the behavior crosses over to a more complicated dependence at higher allowed errors. This transition is likely explained by the different families of measurement bitstrings that satisfy the post-selection criterion in each regime, for example as distinguished by the number of erroneous bits $N_e$, together with their relative combinatorial weights.

Repeated measurements mitigate the detrimental effects of classical readout bit flips. In particular, the resulting survival probability can be substantially larger than the single-shot readout visibility, $1-p_f$. For example, at $\Delta = 0.32$, the majority-vote policy, which has unit survival probability, achieves a measurement error below $10^{-2}$ even for $p_f = 0.08$.

The finite-energy parameter $\Delta$ plays a key role in the overall performance. First, lower values of $\Delta$ reduce the best achievable measurement error by more than an order of magnitude. Second, for a fixed target error, reducing $\Delta$ significantly improves the achievable survival probability. This improvement is consistent with the expectation that lowering $\Delta$ both reduces the single-round measurement error~\cite{hastrup_improved_2021} and suppresses measurement back-action. At the level of joint probability between measurement outcomes, the measurement back-action manifests as correlation in two related ways: First, even a correct measurement leads to a slightly reduced probability of obtaining a correct outcome in subsequent rounds. Second, an incorrect measurement leads to a strong increase in the probability of subsequent errors. We identify this second effect as the primary motivation for allowing our post-selection policies to depend on $N_{fa}$. As a result, the finite-energy parameter $\Delta$, through the associated measurement back-action, sets the main limit on the useful number of repeated measurement rounds.

\bibliography{references_v2}

\end{document}